\documentclass[fleqn,twocolumn,times,baselineskip]{aastex63}
\usepackage{mathtools}

\usepackage{float}
\usepackage{listings}
\usepackage{makeidx}
\usepackage{xcolor}
\usepackage[T1]{fontenc}
\usepackage{braket}
\usepackage{comment}

\accepted{July 18 2021}

\begin{document}
\title{The bias from hydrodynamic simulations: mapping baryon physics onto dark matter fields}

\correspondingauthor{A. Balaguera-Antol\'{\i}nez}
\email{balaguera@iac.es}

\author[0000-0002-0639-8043]{Francesco Sinigaglia}
\affiliation{Department of Physics and Astronomy, Università degli Studi di Padova, Vicolo dell’Osservatorio 3, I-35122, Padova, Italy}
\affiliation{Instituto de Astrof\'isica de Canarias, s/n, E-38205, La  Laguna, Tenerife, Spain}
\affiliation{Departamento  de  Astrof\'isica, Universidad de La Laguna,  E-38206, La Laguna, Tenerife, Spain}

\author[0000-0002-9994-759X]{Francisco-Shu Kitaura}
\affiliation{Instituto de Astrof\'isica de Canarias, s/n, E-38205, La  Laguna, Tenerife, Spain}
\affiliation{Departamento  de  Astrof\'isica, Universidad de La Laguna,  E-38206, La Laguna, Tenerife, Spain}

\author[0000-0001-5028-3035]{A. Balaguera-Antol\'inez}
\affiliation{Instituto de Astrof\'isica de Canarias, s/n, E-38205, La  Laguna, Tenerife, Spain}
\affiliation{Departamento  de  Astrof\'isica, Universidad de La Laguna,  E-38206, La Laguna, Tenerife, Spain}

\author[0000-0001-7457-8487]{Kentaro Nagamine}
\affiliation{Theoretical Astrophysics, Department of Earth and Space Science, Graduate School of Science, Osaka University, \\
1-1 Machikaneyama, Toyonaka, Osaka 560-0043, Japan}
\affiliation{Kavli-IPMU (WPI), University of Tokyo, 5-1-5 Kashiwanoha, Kashiwa, Chiba, 277-8583, Japan}
\affiliation{Department of Physics \& Astronomy, University of Nevada, Las Vegas, 4505 S. Maryland Pkwy, Las Vegas, NV 89154-4002, USA}

\author[0000-0002-5934-9018]{Metin Ata}
\affiliation{Kavli-IPMU (WPI), University of Tokyo, 5-1-5 Kashiwanoha, Kashiwa, Chiba, 277-8583, Japan}

\author{Ikkoh Shimizu}
\affil{Shikoku Gakuin University, 3-2-1 Bunkyocho, Zentsuji, Kagawa, 765-8505, Japan}

\author{M. S\'anchez-Benavente}
\affiliation{Departamento  de  Astrof\'isica, Universidad de La Laguna,  E-38206, La Laguna, Tenerife, Spain}



\begin{abstract}
This paper investigates the hierarchy of baryon physics assembly  bias relations obtained from state-of-the-art  hydrodynamic simulations with respect to the underlying cosmic web spanned by the dark matter field. 
Using the Bias Assignment Method (\texttt{BAM}) we find that non-local bias plays a central role. We classify the cosmic web based on the invariants of the curvature tensor defined not only by  the gravitational potential, but especially by the overdensity, as small scale clustering becomes important in this context. 
First, the gas density bias relation can be directly mapped onto the dark matter density field to high precision exploiting the strong correlation between them. In a second step, the neutral hydrogen is mapped based on the dark matter  and the  gas density fields. Finally, the temperature is mapped based on the previous quantities.
This permits us to statistically reconstruct the baryon properties within the same simulated volume finding percent-precision in the two-point statistics and compatible results in the three-point statistics, in general within 1-$\sigma$, with respect to the reference simulation (with 5 to 6 orders of magnitude less computing time).
This paves the path to establish the best set-up for the construction of mocks probing the intergalactic medium 
for the generation of such key ingredients in the statistical analysis of large forthcoming missions such as DESI, Euclid, J-PAS and WEAVE.
\end{abstract}


% ====================================================================================================
% ====================================================================================================

\section{Introduction} \label{sec:intro}
%We need mocks. We need hydro simulations. We use BAM.
Current and forthcoming galaxy redshift surveys such as DESI \citep{Levi2013}, Euclid \citep{Amendola2018}, J-PAS \citep{Benitez2014} will probe cosmological volumes with an unprecedented number of target galaxies, galaxy clusters, along with a detailed mapping of the inter-galactic medium. Such campaigns will generate three-dimensional maps of the observable Universe, with cosmological information expected to shed light onto profound questions such as the origin of dark energy and the validity of General Relativity as the theory shaping the large-scale structure of the Universe \cite[]{Springel06}. To properly mine such amount of information, robust measurements of cosmological observables and the assessment of their uncertainties (covariance matrices) are unavoidable. The generation of mock galaxy catalogs based on $N$-body simulations has been the standard procedure to obtain covariance matrices. However, for large cosmological volumes and high mass resolution, the construction of thousands of simulated universes is a demanding task both in terms of computing power and memory requirements \citep[][]{Blot2019, Colavincenzo2019, Lippich2019}. 
The situation is much more complex when not only large-scale signals are to  be explored. The resolution of forthcoming experiments promise to wander through small-scale physics, where baryonic phenomena (galaxy formation, feedback by supernovae) can generate a non-negligible footprint in dark matter (DM hereafter) power-spectrum  and the abundance and clustering of DM tracers \citep[see e.g.][]{2008ApJ...672...19R,2013JCAP...04..022B,2015JCAP...12..049S,2015MNRAS.452.2247V,2016MNRAS.456.2361B,2018MNRAS.480.3962C,2019JCAP...03..020S}. This generates the need to account for such effects within the statistical analysis of large redshift galaxy surveys, a task only possible through hydro-simulations, increasing furthermore the degree of complexity, time and memory requirements requested for a robust estimate of covariance matrices in the wake of the era of precision cosmology.
In recent years, a number of alternative approaches to a fast generation of galaxy mock catalogs have seen the light of day. Among these, those based on the so-called bias mapping technique such as \texttt{Patchy} \citep[][]{2014MNRAS.439L..21K} and  \texttt{EZmocks} \citep[][]{2015MNRAS.446.2621C}, have shown
%\citet[][]{2019MNRAS.482.1786L, 2018MNRAS.tmp.2818C,2019MNRAS.485.2806B}
to deliver precise mock catalogs for redshift surveys \citep[][]{Kitaura2016, Zhao2020}. 
All of these methods have been \emph{trained} using $N$-body simulations, from where the halo bias (or the galaxy bias) is calibrated and used to sample independent sets of dark matter density fields (DMDF hereafter)

The Bias Assignment Method  \citep[\texttt{BAM} hereafter,][]{Balaguera2018} represents the latest attempt to use the concept of bias-mapping to generate mock catalogs. When applied to tracers such as dark matter haloes, this method has been successfully shown to generate $~2\%-5\%$ precision in the two (e.g power-spectrum) and three-point (e.g. bi-spectrum) statistics of such tracers, with similar figures for the accuracy of the corresponding covariance matrices \cite[][]{Balaguera2019}.
With this motivation, in this article we extend the \texttt{BAM} approach into the context of hydro-simulations, aiming at establishing the best strategy to reproduce, to percent precision the statistical distribution (up to the three-point statistics) of different baryon properties (e.g. gas density, gas temperature, density of neutral hydrogen). Such strategies are planned to be used in the generation of mock catalogs for experiments focused on the properties and cosmological content of the  inter-galactic medium. 

To put the present work into context we should mention a few recent developments. \citet[][]{2017MNRAS.470..340P} and \citet{2017MNRAS.469.2323P}
presented the analysis of abundance and clustering of neutral hydrogen based on a halo-model framework.
\citet{2017MNRAS.471.1788C} and \citet[][]{VillaescusaNavarro2018} extended that application and explored the possibility of simulating neutral hydrogen (HI) intensity maps applying the concepts of halo-occupation distribution and perturbative methods to distribute HI inside dark matter halos \citep[see also][for an extension of this approach]{Ando20}. All these methods allow for the construction of mock 21cm maps by relying on dark matter-only $N$-body simulations, which are less expensive than the full hydrodynamic approach and can be carried out on larger volumes. Even though promising and physically well-motivated, the method throws deviations in the two-point statistics of $\gtrsim5\%$ with respect to a full hydro-simulation. A similar goal has been recently pursued by \citet[][]{Dai2020}, who combined the \texttt{FastPM} \citep{2016MNRAS.463.2273F} with a deep-learning approach in order to generate simulations with baryonic processes such as cooling, radiation, star formation, gas shocks and turbulence. This approach requires a cosmological hydrodynamic simulation from which a number of parameter is extracted, leading to reconstructions displaying power spectra with differences of the order of $>10\%$ with respect to the reference. 

These studies (including the present work) represent independent attempts to produce cosmological volumes with baryonic physics, without the need of running full hydrodynamic simulations. The unique feature of the approach presented in this paper is that our methodology is not a machine-learning oriented one. Instead, it relies on a explicit description of the link (or bias) between the statistical distribution of baryonic properties and the underlying DMDF, extracted from one reference hydro-simulation. The precision of the reconstruction of the statistical properties of the baryon distribution (two and three-point statistics) depends on the physical information extracted from properties of the DMDF (such as the tidal field or the distribution of peaks), thus providing a physical framework within which the different outcomes of our reconstruction procedure can be interpreted.

The paper is organized as follows. In \S\ref{sec:bam} we briefly introduce the \texttt{BAM} method. In \S\ref{sec:refsim} we describe the reference cosmological hydrodynamic simulation. In \S\ref{sec:hydro_reproduction} we describe the reconstruction for the \texttt{BAM} approach and end with conclusions in \S\ref{sec:conclusions}.

% ====================================================================================================
% ====================================================================================================

\section{BAM: The Bias Assignment Method} \label{sec:bam}
\texttt{BAM}  \citep[][BAM-I and BAM-II hereafter]{Balaguera2018,Balaguera2019} represents the most recent version of a bias-mapping technique. The method aims at generating mock catalogs of DM tracers in a precise and efficient fashion. The main concept behind this approach is a learning process, based on one single reference (dark-matter only or hydro) simulation, of the tracer-dark matter bias relation. This relation (understood as the probability of having a given number of DM tracers conditional to a set of properties of the DMDF), can be used to generate independent realizations of the tracer spatial distribution, when mapped over independent DMDF (evolved from a set of initial conditions generated the cosmological model and parameters used for the reference simulation). \texttt{BAM} has been shown to be able to generate ensembles of mock catalogs with an accuracy of $1-10\%$ in the two- and three-point statistics of dark matter halos  \citep[e.g.][]{Pellejero2020} and currently aims at generating the realistic set of galaxy mock catalogs by including intrinsic properties of tracers such as halo masses, concentrations, spins (Balaguera-Antol\'{\i}nez \& Kitaura, in preparation).
%%%%%%%%%%%%%%%%%%%%%%%%%%%%%%%%%%%%%%%%%%%%%%%%%%%%%%%
\begin{table}
    \centering
    \begin{tabular}{ccccc}
    \toprule
  Cosmic Web Type &VF & $\langle \rho_{\rm gas}\rangle$ & $\langle T_{\rm gas}\rangle$ & $\langle n_{HI}\rangle$ \\
    \toprule
  All  & $100\%$ &$1.1$ & $4.1\times 10^{4}$ & $0.3$  \\
  Knots  & $1.9\%$ &$9.1$ & $3.3\times 10^{5}$ & $9.7$  \\
  Filaments  &$27.1\%$ &$1.9$ & $8.9\times 10^{4}$ & $0.4$   \\
  Sheets  &$55.5\%$ &$0.6$ & $1.7\times 10^{4}$ & $0.01$  \\
  Voids  &$15.5\%$ &$0.2$ & $5.9\times 10^{3}$ & $10^{-4}$  \\
    \toprule
    \end{tabular}
    \caption{\small{Mean values of baryon properties in the hydrodynamic simulation used in this work. The cosmic-web classification is defined in \S~\ref{sec:bias}. $VF$ stands for the volume fraction occupied by of each cosmic-web type in the simulation. The gas density is expressed in units of $10^{-29}{\rm gr}/{\rm cm}^{3}$, the gas temperature in Kelvins and the number density of HI in $10^{-7}{\rm cm}^{-3}$.}}
    \label{table:gas_prop}
\end{table}
%%%%%%%%%%%%%%%%%%%%%%%%%%%%%%%%%%%%%%%%%%%%%%%%%%%%%%%
\begin{figure}
\includegraphics[width=8.5cm]{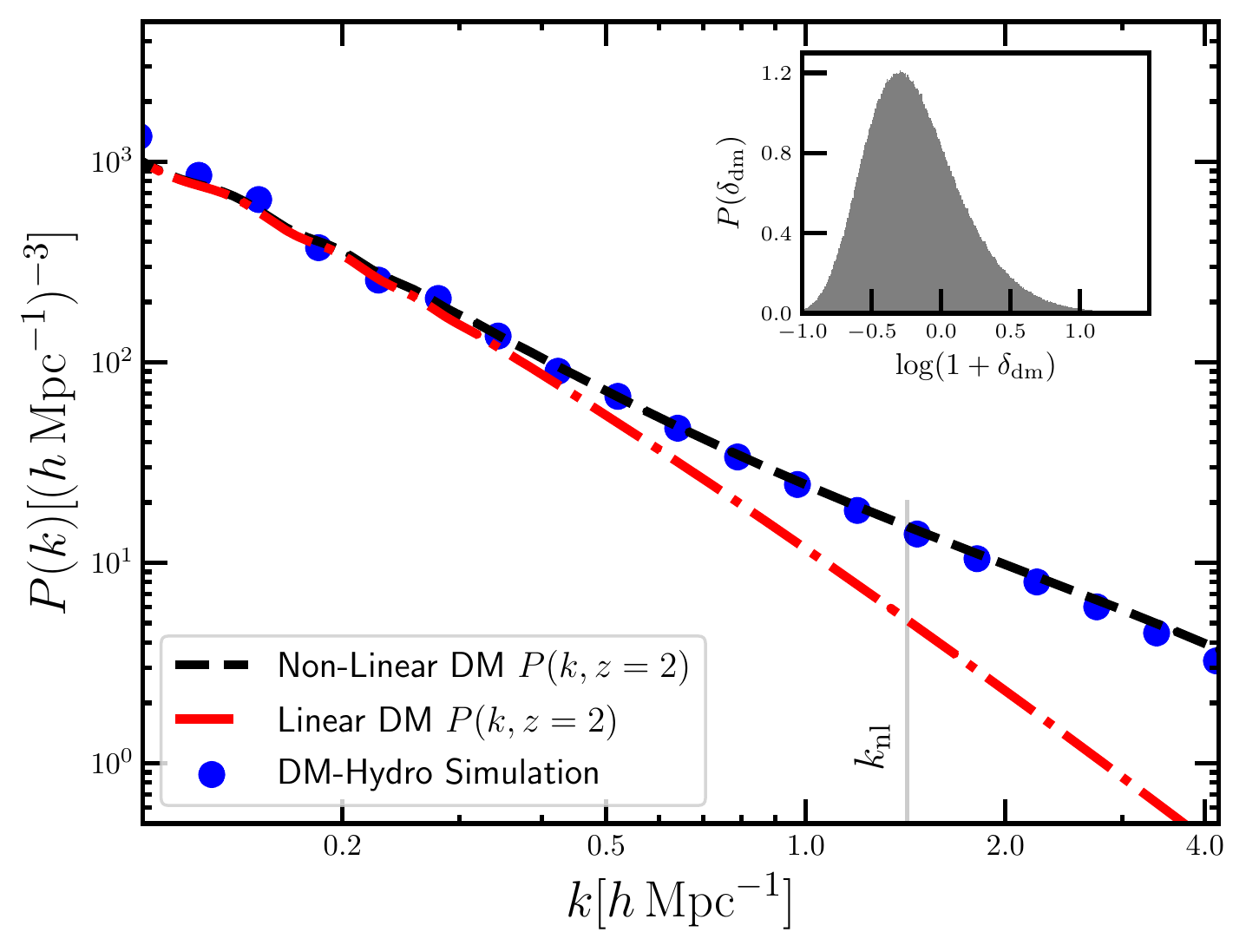}
\caption{\small{power-spectrum  of the dark matter distribution in the Hydro-simulation (solid line, measured on a $512^{3}$ mesh). The dotted and dashed lines show respectively the predictions of linear and non linear (\texttt{Halo-Fit}) matter power-spectrum  at $z=2$. The vertical line denotes the scale $k_{\rm nl}\sim 1.4\,h$ Mpc$^{-1}$ at which the r.m.s of matter fluctuations $\sigma(k)$ equals unity. The inset plot shows the distribution of the dark-matter overdensities in the hydro-simulation.}}
\label{fig:dm_power}
\end{figure}
%%%%%%%%%%%%%%%%%%%%%%%%%%%%%%%%%%%%%%%%%%%%%%%%%%%%%%%

In this work we extend the approach of \texttt{BAM} to the context of hydrodynamic simulations. Provided a reference hydro-simulation, we aim at statistically reproducing the one-, two- and three-point statistics of the spatial distribution of properties of the simulated gas (ionized gas density, abundance of neutral hydrogen, gas temperature), through the assessment of the bias relation between the distribution of different gas properties and the underlying DMDF. This quantity (the bias) is represented by a multi-dimensional probability distribution of having a certain gas property $\eta$ (e.g., $\eta=$ gas density, gas temperature, number density of neutral hydrogen) interpolated on a cell of volume $\partial V$ conditional to the properties of the underlying DMDF ${\Theta}$ at the same cell, $\mathcal{P}(\eta| {\Theta})_{\partial V}$ (we will omit the symbol $\partial V$ hereafter). The statistical nature of this bias implies that not only average trends in the link between cosmological and astrophysical quantities are taken into account, but also the intrinsic scatter among these.

%%%%%%%%%%%%%%%%%%%%%%%%%%%%%%%%%%%%%%%%%%%%%%%%%%%%%%%
\begin{figure*}
\hspace{-1.2cm}
\includegraphics[width=19.cm,height=7cm]{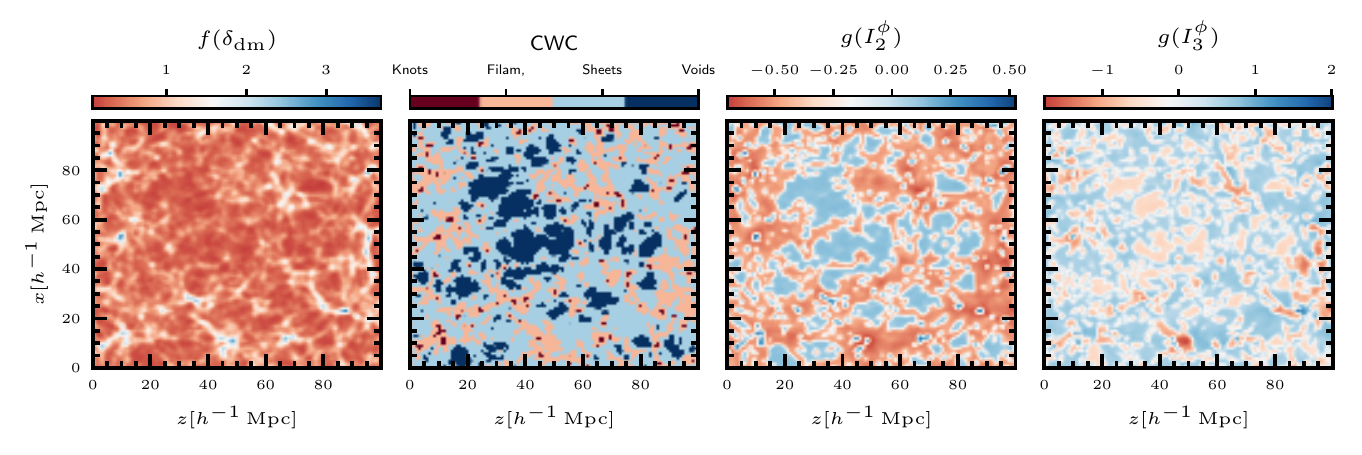}
\centering
\includegraphics[width=15cm, height=7cm]{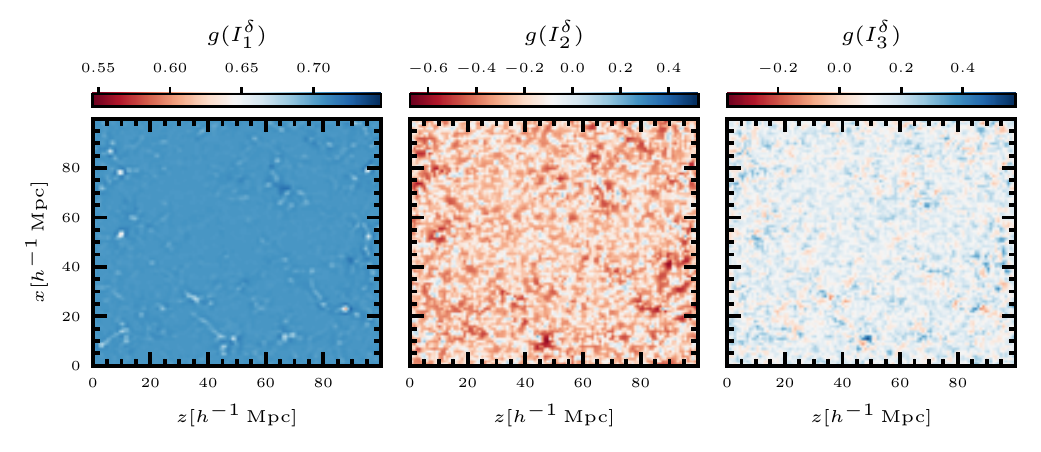}
\caption{\small{Slices of $\sim 10h^{-1}\,$Mpc thickness from the volume of the reference simulation showing different properties of the dark matter density field transformed according the functions $f,g(x)$ (see \S~\ref{sec:bias}). First (left-to-right) panel shows dark matter distribution. Second panel shows the cosmic-web classification. The third and fourth panel show the second and third invariant $g(I_{2}), g(I_{3})$, of the tidal field. Bottom panels show the terms $g(I^{\delta}_{i})$ where $I^{\delta}_{1}=\nabla^{2}\delta_{\rm dm}$.}}
\label{fig:slices_dm_prop}
\end{figure*}
%%%%%%%%%%%%%%%%%%%%%%%%%%%%%%%%%%%%%%%%%%%%%%%%%%%%%%%

The \texttt{BAM} procedure can be briefly summarized as follows:
\begin{itemize}
    \item Start with a \texttt{DMDF} $\delta_{\rm dm}(\vec{r})$ interpolated on a mesh with resolution $\partial V$.
    \item For a given gas property $\eta$, obtain the interpolation on the same mesh as the DMDF. 
    \item Measure the bias $\mathcal{P}(\eta|\{\Theta\})$, where $\{\Theta\}$ denotes a set of properties (presented in \S \ref{sec:bias}) of the underlying dark matter density field $\delta_{\rm dm}(\vec{r})$ retrieved from  the reference simulation. This bias is obtained from the assessment of the joint distribution $\mathcal{P}(\eta,\{\Theta\})$, which is represented by a multi-dimensional histogram consisting on $N_{\eta}$ bins of the target tracer property and $N_{\Theta}$ properties of the DMDF.
    
    \item Sample a new variable $\tilde{\eta}$ at each spatial cell (with DM properties $\{\tilde{\Theta}\}$) as 
    \begin{equation}\label{eq:sam_raw}
        \tilde{\eta}\curvearrowleft \mathcal{P}(\tilde{\eta}=\eta|\{\Theta\}=\{\tilde{\Theta}\}) 
        \end{equation}
    \item Measure the power-spectrum  of this quantity and compare with the same statistics from the reference.
\end{itemize}
As shown in BAM-I, this procedure leads to a power-spectrum  of the variable $\tilde{\eta}$, $P_{\tilde{\eta}}(k)$, which is biased with respect to the reference statistics $P_{\eta}(k)$. This difference is inversely proportional to the amount of information encoded in $\{\Theta\}$. In order to compensate for the missing physical information, \texttt{BAM} performs an iterative process in which the DM density field (and hence its properties) is updated through the convolution with an isotropic kernel $\mathcal{K}(k=|\vec{k}|)$. Briefly, the process to obtain the kernel starts with the definition of the ratio $T_{0}(k)\equiv P^{i=0}_{\tilde{\eta}}(k)/P_{\eta}(k)$, where $P^{i=0}_{\tilde{\eta}}(k)$ is the power-spectrum  obtained from the sampling described by Eq.~(\ref{eq:sam_raw}). At each iteration $i$ (producing a new estimate of $P^{i}_{\tilde{\eta}}(k)$), a value of the ratio $T_{i}(k)$ is computed. At each wavenumber $k$, it is subject to a Metropolis-Hasting (MH) rejection algorithm\footnote{This is done by computing a transition probability ${\rm min}(1, {\rm exp}(\mathcal{H}^{2}_{0j}-\mathcal{H}^{2}_{1j}))$ with $\mathcal{H}_{ij}=(P_{\rm ref}-P^{i}_{\tilde{\eta}})/\sigma$, $\sigma=\sigma(k)$ the (Gaussian) variance associated to the reference power-spectrum.}. The outcome of the MH sampling process defines a sets of weights $w_{i}(k)=T_{i}(k)$ if the current value of $T_{i}(k)$ is accepted, or $w_{i}(k)=1$ otherwise. The \texttt{BAM} kernel is then defined as $\mathcal{K}(k)\equiv \prod_{j=0}^{j=i}w_{j}(k)$ such that, under successive convolutions with the DMDF, the limit $\lim_{n\to \infty} P^{(n)}_{\tilde{\eta}}(k)/P_{\eta}(k)\to 1$ (where $n$ is the number of iterations) is induced. 
In practice, the iterative process is stopped when the residuals\footnote{Throughout this work, residuals between two quantities $A(x)$ and $B(x)$ are computed as $\sum_{i}|A(x_{i})/B(x_{i})-1|/N_{x}$ where the sum is done over all probed (bins of) values of the variable $x$ and $N_{x}$ is the number of $x$ values.} between these power spectra are of the order of $1\%$, which for the current set-up is reached with $n \sim 200$ iterations.

We stress that the iterative process within BAM is not guaranteed to achieve convergence by its own. The degree of convergence is highly sensitive to i) the cross-correlation between the quantity to be reconstructed and the DMDF, ii) the choice of properties of the DMDF and iii) the way these are mined (i.e, binning). A weak fulfilment of any of these requirements can go in detriment of the convergence of the method, as  is indeed the case of the gas temperature, as will be exposed below.

As will be explained below (section \S\ref{sec:bias}, see BAM-I and BAM-II), we divide the set of properties of the DMDF in local (e.g., density) and non-local (e.g., tidal field). Such classification can be linked to coefficients in a perturbative expansion of the DMDF involving local (e.g., $\delta, \delta^{2}$) and non-local quantities (i.e., the tidal field), as explained by \citet[][]{Kitaura2020}. Furthermore, the selection of a given set $\{\Theta\}$ can be associated to a particular choice of terms in such perturbative expansion, thus neglecting sources of local and non-local information (e.g., peak distribution, velocity shear). The \texttt{BAM}-kernel can be then understood as accounting for the contributions of such missing terms. Even though the convolution with the local density will generate a non-local component such that, within the iterative process all quantities composing the set $\{\Theta\}$ are strictly speaking non-local, the amount of such induced non-local behavior depends on the shape of the kernel. As an example, a constant kernel $\mathcal{K}(k)$ represents a re-normalization of the DM fluctuations on all scales, hence in practice not inducing non-local information. Such scenario would imply that the missing information in the assessment of the tracer bias is indeed of local nature. A scale-dependent kernel, on the contrary, induces non-local behavior, though it is not evident that the missing information in that case is purely non-local. Furthermore, such non-local behavior contains also the effects of aliasing inherently present due to the implementation of a mass assignment scheme in order to interpolate the different gas properties.

As we will show in \S\ref{sec:hydro_reproduction}, the kernels obtained under different choices of $\{\Theta\}$, when applied to the reconstruction of the gas density and the neutral hydrogen, display a constant behavior on large scales and scale dependencies on small scales, suggesting that non-locality induced within the iterative process through the convolution of the local density (which mainly shapes the large scale signal of biased tracers) is sub-dominant compared to the originally defined non-local quantities.

%%%%%%%%%%%%%%%%%%%%%%%%%%%%%%%%%%%%%%%%%%%%%%%%%%%%%%%
\begin{figure*}
\hspace{-1.5cm}
\includegraphics[width=21cm]{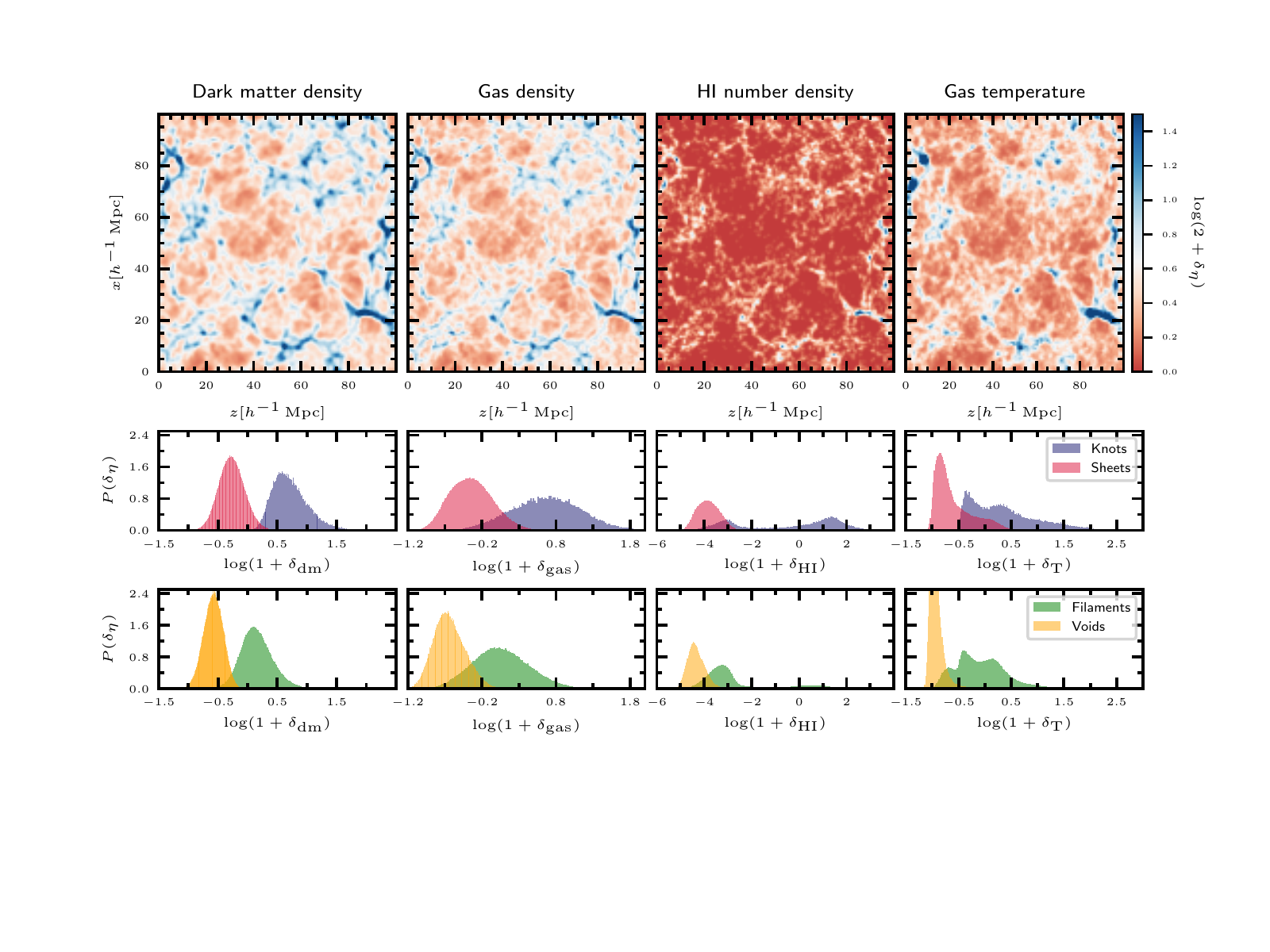}
\vspace{-4cm}
\caption{\small{Top-row: slices of $\sim 10h^{-1}$\,Mpc thick from the volume of the reference simulation, showing the spatial distribution of dark matter, the ionized gas density, HI number density and ionized gas temperature. The color scale is the same for all four panels,  and denotes values of $\log(2+\delta_{\eta})$, where $\eta$ refers to the overdensity of the different properties. Bottom rows: distribution of dark matter and gas properties in different cosmic-web types. The cosmic-web classification is obtained from the overdensity field of the dark matter distribution (see \S\ref{sec:bias}).}}
\label{fig:slices_ref_comp}
\end{figure*}
%%%%%%%%%%%%%%%%%%%%%%%%%%%%%%%%%%%%%%%%%%%%%%%%%%%%%%%
In summary, the result of the iterative procedure is composed of three outputs. First, the interpolated $\tilde{\eta}$-field, which shares the same one- and two-point statistics as the reference (the precision in the three-point statistics of the new $\tilde{\eta}$-field depends on the characterization of the set of properties $\{\Theta\}$, as we will discuss in \S~\ref{sec:bias}). 
Second, the isotropic kernel $\mathcal{K}(k)$ and third, a \texttt{BAM}-bias $\mathcal{P}(\eta|\hat{\Theta})$, where $\{\hat{\Theta}\} \equiv \mathcal{K}\otimes \{\Theta\}$ \footnote{With this notation we indicate that the convolution acts over the density field $\delta_{\rm dm}$ from which the set of properties $\Theta$ are determined.}. The latter two quantities are meant to be used onto a new realisation of the DMDF (with properties $\{\Theta\}_{\rm new}$) to generate a new realisation of the $\eta$ field as
\begin{equation}\label{eq:sam_mock}
        \eta_{\rm new}\curvearrowleft \mathcal{P}(\eta_{\rm new}=\eta| \mathcal{K}\otimes \{\Theta\}_{\rm new}=\{\tilde{\Theta}\}). 
\end{equation}
In this work, we assess the best strategy with which the different set of DM properties can be used to obtain accurate summary statistics from the different gas properties in the reference hydro-simulation. In forthcoming publications we will show the application of these products and strategies to independent density fields. Building on the findings of BAM-II, we will show in future work that the iterative procedure does not fall into overfitting when applied to independent dark matter density fields. The assessment of this feature will pave the path towards the generation of mock catalogs with applications in Lyman-$\alpha$ and $21$-cm intensity mapping surveys.

%%%%%%%%%%%%%%%%%%%%%%%%%%%%%%%%%%%%%%%%%%%%%%%%%%%%%%%
\begin{figure*}
\hspace{-1.5cm}
\includegraphics[width=21cm]{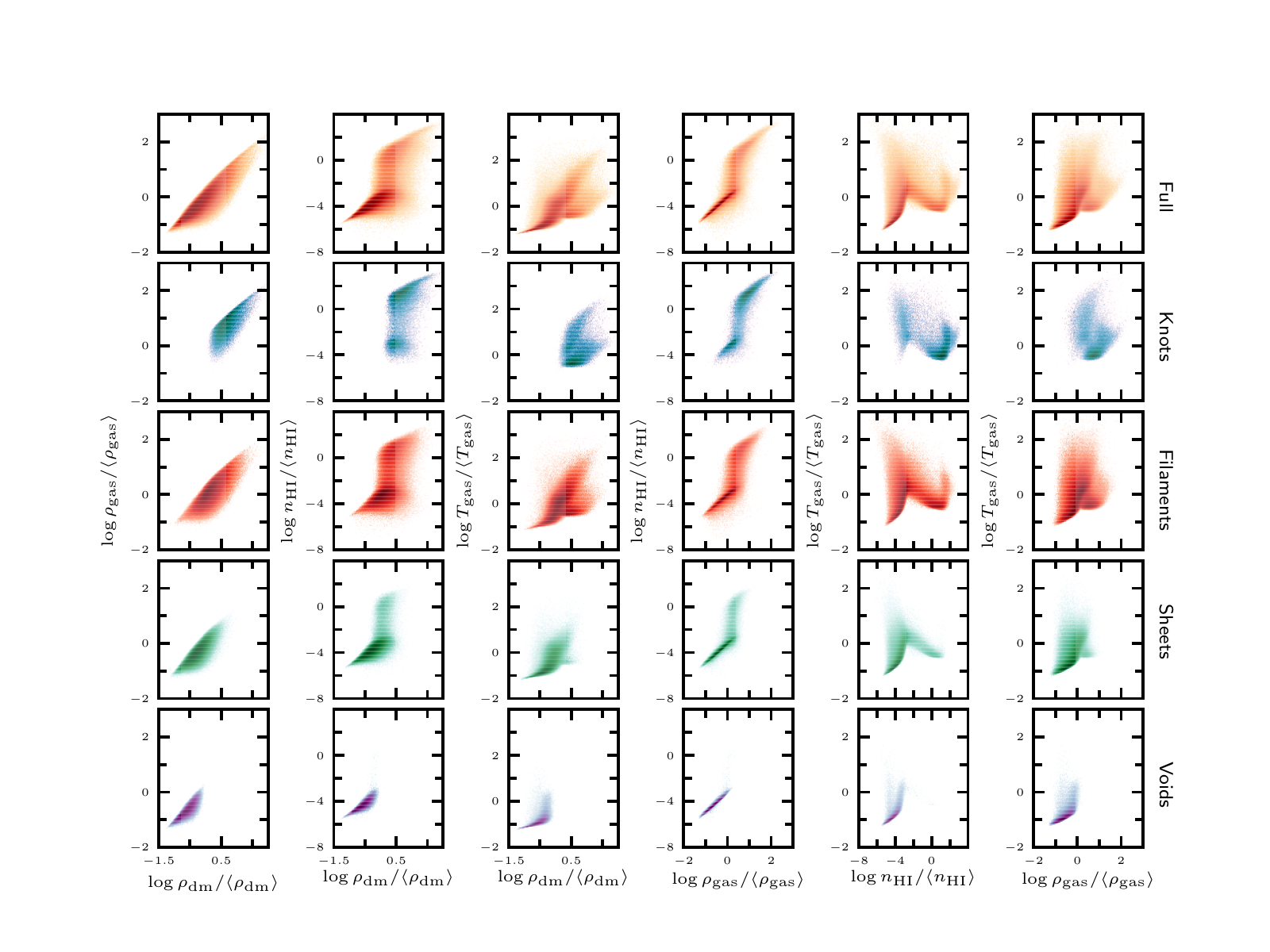}
\vspace{-1.2cm}
\caption{\small{Phase-space diagrams of the simulated gas identified within different cosmic-web types (see \S~\ref{sec:bias}). The color intensity indicates the height of the joint probability distribution between each pair of properties.}}
\label{fig:hydro_cw}
\end{figure*}
%%%%%%%%%%%%%%%%%%%%%%%%%%%%%%%%%%%%%%%%%%%%%%%%%%%%%%%

% ====================================================================================================
% ====================================================================================================

\section{The hydro-simulation} 
\label{sec:refsim}
The reference simulation has been run with the cosmological smoothed-particle hydrodynamics (SPH) code \texttt{Gadget3-OSAKA} \citep{Aoyama2018, Shimizu2019}, a modified version of the popular $N$-body/SPH code \texttt{Gadget2} \citep{Springel2005}. It embeds a comoving volume $V=(100h^{-1}\text{Mpc})^3$, large enough to perform restricted cosmological studies, and $N=2\times512^3$ particles of mass $m_{\rm DM}=5.38\times10^8h^{-1}\text{M}_\odot$ for DM particles and $m_{\rm gas}=1.0\times 10^8 h^{-1}\text{M}_\odot$ for gas particles.
\footnote{For comparison, our resolution is comparable to that of the Illustris-3 simulation (\url{https://www.illustris-project.org/data/}).} 
%the Ilustris simulation \cite[e.g.][]{2014MNRAS.444.1518V} displays two orders of magnitude higher in the particle mass in approximately the same cosmological volume.} % and gravitational softening length of $6h^{-1}$kpc (comoving). 
The gravitational softening length is set to $\epsilon_g = 7.8 h^{-1}$ kpc (comoving), but we allow the baryonic smoothing length to become as small as $0.1\epsilon_g$. This means that the minimum baryonic smoothing at $z=2$ is about physical $260\,h^{-1}$\,pc, which is sufficient to resolve the HI distribution in the circumgalactic medium. 
The star formation and supernova feedback are treated as described in \citet{Shimizu2019}.
We note that the same simulation has been used for the Ly$\alpha$ forest analyses \citep{Momose20a,Nagamine20}, and the same code has been used for the study of cosmological HI distribution \citep{Ando19}.
The code contains also important refinements, such as e.g. the density-independent formulation of SPH and the time-step limiter \citep[][]{Hopkins2013, Saitoh2013}. 

The main baryonic processes which shape the evolution of the gas are photo-heating, photo-ionization under the UV background \cite[][]{Haardt2012} and radiative cooling. All these processes are accounted for and solved by the \texttt{Grackle} library \citep{Smith2017}, which determines the chemistry for atomic (H, D and He) and molecular (H$_2$ and HD) species. The initial conditions are generated at redshift $z=99$ using \texttt{MUSIC} \citep{Hahn2011} with cosmological parameters taken from \cite{Planck2016}. 
In this work we use the output at $z=2$ (for which the computation times amounts to $\sim 1,04 \times 10^{5}$ CPU hours), reading gas properties such as the ionized gas density, the HI number density and the gas temperature. 
Table~\ref{table:gas_prop} shows the mean values of these properties within the simulated volume, as well as in different cosmic-web types (defined in \S~\ref{sec:bias}).

%%%%%%%%%%%%%%%%%%%%%%%%%%%%%%%%%%%%%%%%%%%%%%%%%%%%%%%
\begin{figure*}
\includegraphics[width=17cm]{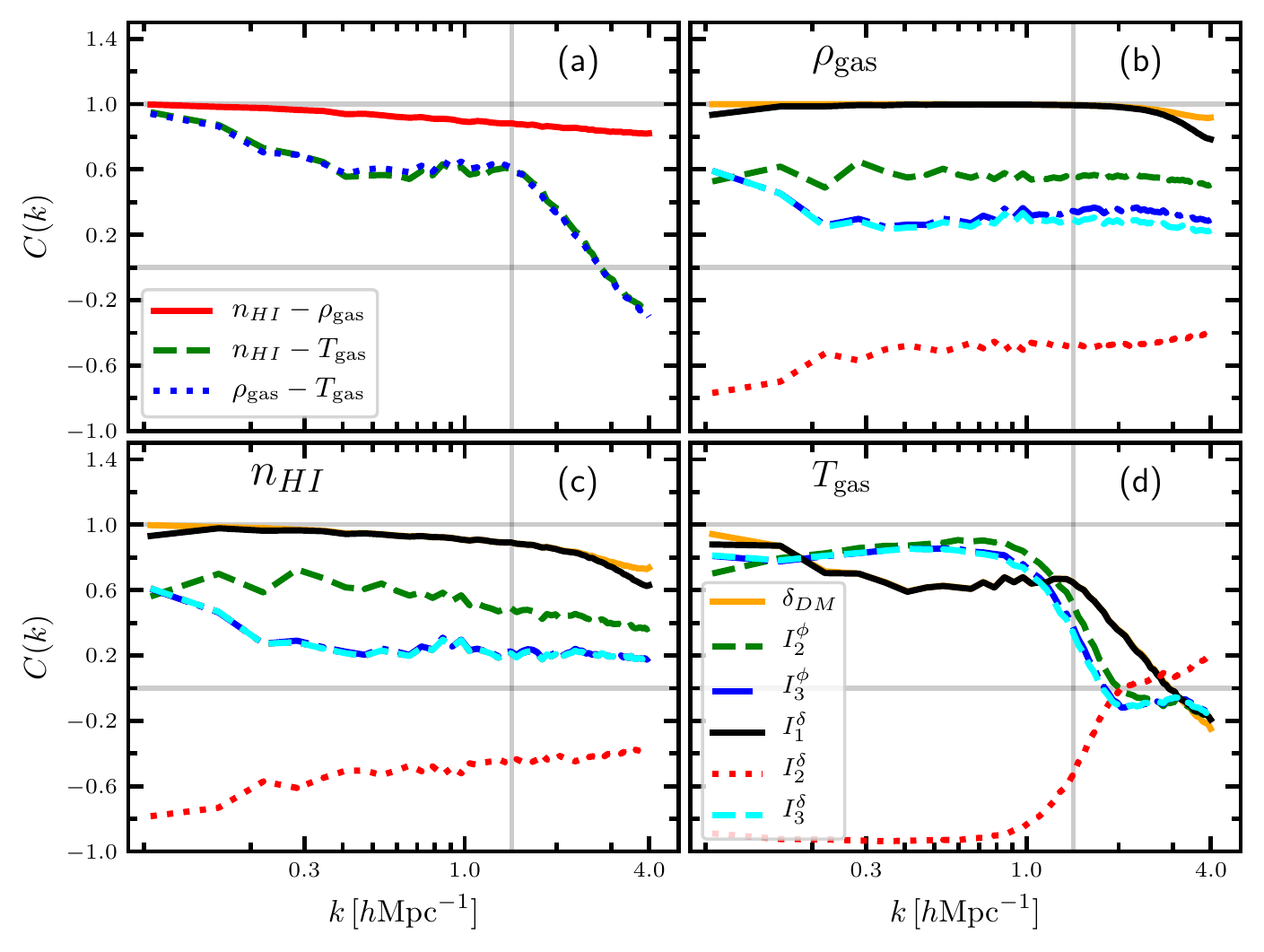}
\caption{\small{Correlation coefficients in Fourier space among the different baryon properties (panel (a)) and with the  different properties inferred from the DMDF (panels (b) to (d)). The horizontal grey lines mark the limits $C(k)=1$ and $C(k)=0$. The vertical line corresponds to the value $k_{\rm nl}=1.4h\,$ Mpc$^{-1}$ (see \S\ref{sec:bias}) denoting the transition from linear ($k\leq k_{\rm nl})$ to non-linear ($k>k_{\rm nl}$) scales. The legend in panel (d) applies for panels (b) and (c) as well.}}
\label{fig:cross_corr_refs}
\end{figure*}
%%%%%%%%%%%%%%%%%%%%%%%%%%%%%%%%%%%%%%%%%%%%%%%%%%%%%%%

These properties, as for the DMDF, have been interpolated into a $128^3$ cubic mesh using a CIC mass assignment scheme (MAS hereafter). Such resolution corresponds to a physical cell- volume of $\partial V \sim (0.8h^{-1}\rm{Mpc})^{3}$ and a Nyquist frequency of $k_{N}=4.02\,h\,$Mpc$^{-1}$. The spherical averages in Fourier space done to obtain power-  and bi-spectrum are performed within shells of width $\Delta k=2\pi/V^{1/3}$, i.e., the fundamental mode of the simulated volume. The measurements of power spectra and bi-spectra are corrected using the Fourier transform of the MAS implemented \citep[see e.g.][]{1981csup.book.....H} as well as with a Poisson model for their respective shot-noise subtraction.

%%%%%%%%%%%%%%%%%%%%%%%%%%%%%%%%%%%%%%%%%%%%%%%%%%%%%%%
\begin{figure*}
\hspace{-1cm}
\includegraphics[width=20cm]{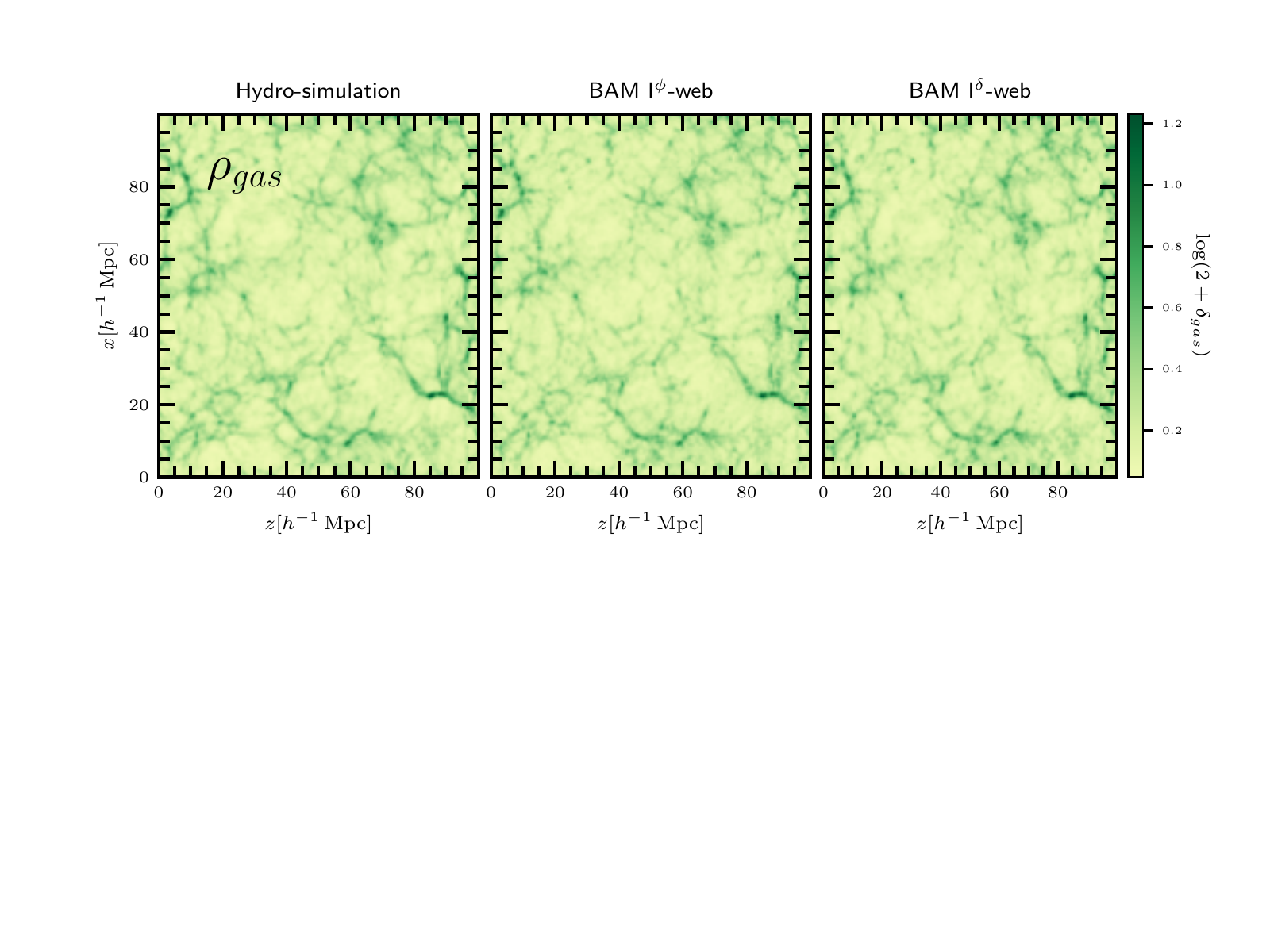}
\vspace{-7cm}
\caption{\small{Slices of $\sim 10\,h^{-1}\,$Mpc thick of ionized gas density obtained from the reference simulation (top-left panel) and the \texttt{BAM} calibration procedure using different set of DM properties.}} 
\label{fig:slices_gas}
\end{figure*}
%%%%%%%%%%%%%%%%%%%%%%%%%%%%%%%%%%%%%%%%%%%%%%%%%%%%%%%

The evolution of baryons within a hydrodynamic simulation leaves imprints on the statistics of the dark matter distribution and its tracers, with sizable differences, on small scales, when compared dark-matter only simulations based on the same initial conditions \citep[see e.g.][]{2008ApJ...672...19R,2012MNRAS.423.2279C,2015JCAP...12..049S,2013JCAP...04..022B, 2014MNRAS.440.2290M,2015MNRAS.452.2247V,2016MNRAS.456.2361B,2018MNRAS.480.3962C}. In general, in the presence of this type of back-reaction, the dark matter distribution (or any of its properties) cannot be properly referred as an independent variable with respect to which all posterior distributions used in \texttt{BAM} are expressed. For the present work, this poses no conflict as long as we will refer to the baryon and DM properties from the same simulation. 
For the purpose of producing mocks based on approximated methods (as will be discussed in forthcoming publications), the lack of baryonic effects in such approximated DMDF can be in principle degenerated with the lack of precision on small scales of such approximated methods, an aspect solved through the iterative procedure and the \texttt{BAM} kernel described in \S\ref{sec:bam}.

For the current hydro-simulation, the back-reaction of dark matter under the presence of baryons is expected to be small. This claim is motivated by Fig.~\ref{fig:dm_power}, where we show the power-spectrum  of the dark matter distribution from the hydro-simulation, together with accurate predictions (using the same set of cosmological parameters) of a linear power-spectrum  \citep[][]{1998ApJ...496..605E} and a non-linear prediction based on fits to dark-matter only simulations \cite[][]{2003MNRAS.341.1311S,2012ApJ...761..152T}. The DM power-spectrum  follows closely the non-linear prediction, well beyond the non-linear scale $k_{\rm nl}$ defined by the condition $\sigma(k=k_{\rm nl})=1$  \footnote{$\sigma^{2}(k_{*},z)=(2\pi^{2})^{-1}\int {\rm d}k \, k^{2}P(k,z)|W(k/k_{*})|^{2}$ is the r.m.s of the mass distribution, $P(k,z)$ is the linear matter power-spectrum  at a given redshift $z$ and $W(k/k_{*})$ is a top-hat filter in Fourier space with characteristic scale $k^{-1}_{*}$.}, in this case $k_{\rm nl}\sim 1.4 h\, {\rm Mpc}^{-1}$.

% ====================================================================================================
\subsection{Properties of the dark matter density field }
\label{sec:bias}
A key step to unleash the \texttt{BAM} machinery is the assessment of the set of properties of the underlying DMDF, which will be used to statistically characterize the abundance of dark matter tracers (gas properties in our case). In general, these properties can be divided in i) local properties (i.e., dark matter density in a cell) and ii) non-local properties (i.e. tidal field). BAM-I and BAM-II explored the zero-th order non-local treatment through the cosmic-web classification, based on the behavior of the eigenvalues of the tidal field tensor $\mathcal{T}_{ij}\equiv \partial_{i}\partial_{j}\,\phi$ \citep[where $\phi$ denotes the gravitational potential, see e.g.][]{Hahn2007,2018MNRAS.473.1195L} with respect to an arbitrary threshold $\lambda_{\rm th}=0$ \citep[see e.g.][]{ForeroRomero2009}. This classification (T-web hereafter) generates knots ($\lambda_{i}> \lambda_{\rm th}\,\forall i$), filaments/sheets (two/one eigenvalues $\lambda_{i}>\lambda_{\rm th}$, one/two $\lambda_{i}<\lambda_{\rm th}$) and voids ($\lambda_{i}< \lambda_{\rm th}\,\forall i$). Furthermore, environmental properties such as the mass of collapsing regions (defined as a friend-of-friend collection of cells classified as knots) was also shown to introduce relevant physical information by improving the precision of the three-point statistics from the \texttt{BAM} reconstructions (see e.g. BAM-I and \cite{2015MNRAS.451.4266Z}).

%%%%%%%%%%%%%%%%%%%%%%%%%%%%%%%%%%%%%%%%%%%%%%%%%%%%%%%
\begin{figure*}
\centering
\includegraphics[width=18cm]{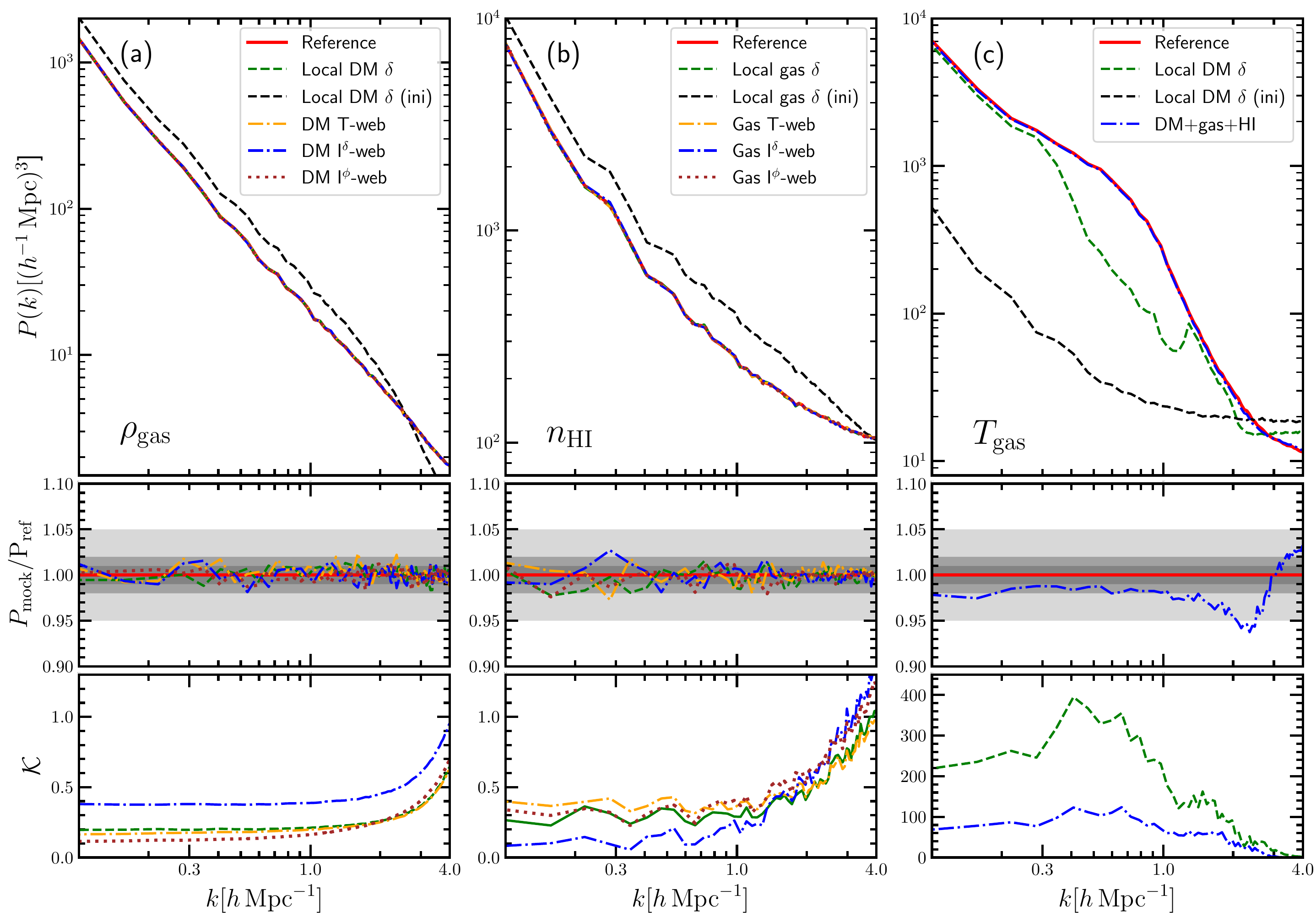}
\caption{\small{Power spectra of ionized gas, neutral Hydrogen and gas temperature as obtained by \texttt{BAM} compared to the same quantities obtained from the reference simulation. Top panels show the results of the reconstruction using different sets of underlying DM properties. These panels also shows the power-spectrum  obtained at the first iteration (ini) in the local-bias model. Middle panels show the ratio of each of power spectra to the measurements obtained from the reference simulation. Gray shaded areas in these panels represent the $1\%,2\%,5\%$ deviations with respect to unity. Bottom panels show the \texttt{BAM} kernel $\mathcal{K}(k)$ resulting from the calibration procedure for each set $\{\Theta\}$ explored.}}
\label{fig:pk_final}
\end{figure*}
%%%%%%%%%%%%%%%%%%%%%%%%%%%%%%%%%%%%%%%%%%%%%%%%%%%%%%%

\cite{Kitaura2020} extended the use of tidal field properties in \texttt{BAM}, and explored the information content in terms of some specific combinations of the eigenvalues $\lambda_{i}$. In particular, that work showed that using the principal invariants of the tidal field $I^{\phi}_1\equiv \lambda_1+\lambda_2+\lambda_3$ (the trace, which coincides with the overdensity $\delta_{\rm dm}$), $I^{\phi}_2\equiv \lambda_1\lambda_2+\lambda_2\lambda_3+\lambda_1\lambda_3$ and $I^{\phi}_3 \equiv \lambda_1\lambda_2\lambda_3$ (the determinant)
 leads to reconstructions with $\sim 5 \%$ precision (with respect to the reference) in the three-point statistics of dark matter halos. Our numerical experiments with hydro-simulation described in \S\ref{sec:refsim} have confirmed that trend, i.e., that the information encoded in the invariants (I$^{\phi}$-web hereafter) is statistically larger than that encoded in the T-web. 
%%%%%%%%%%%%%%%%%%%%%%%%%%%%%%%%%%%%%%%%%%%%%%%%%%%%%%%
\begin{figure}
\includegraphics[width=8.5cm]{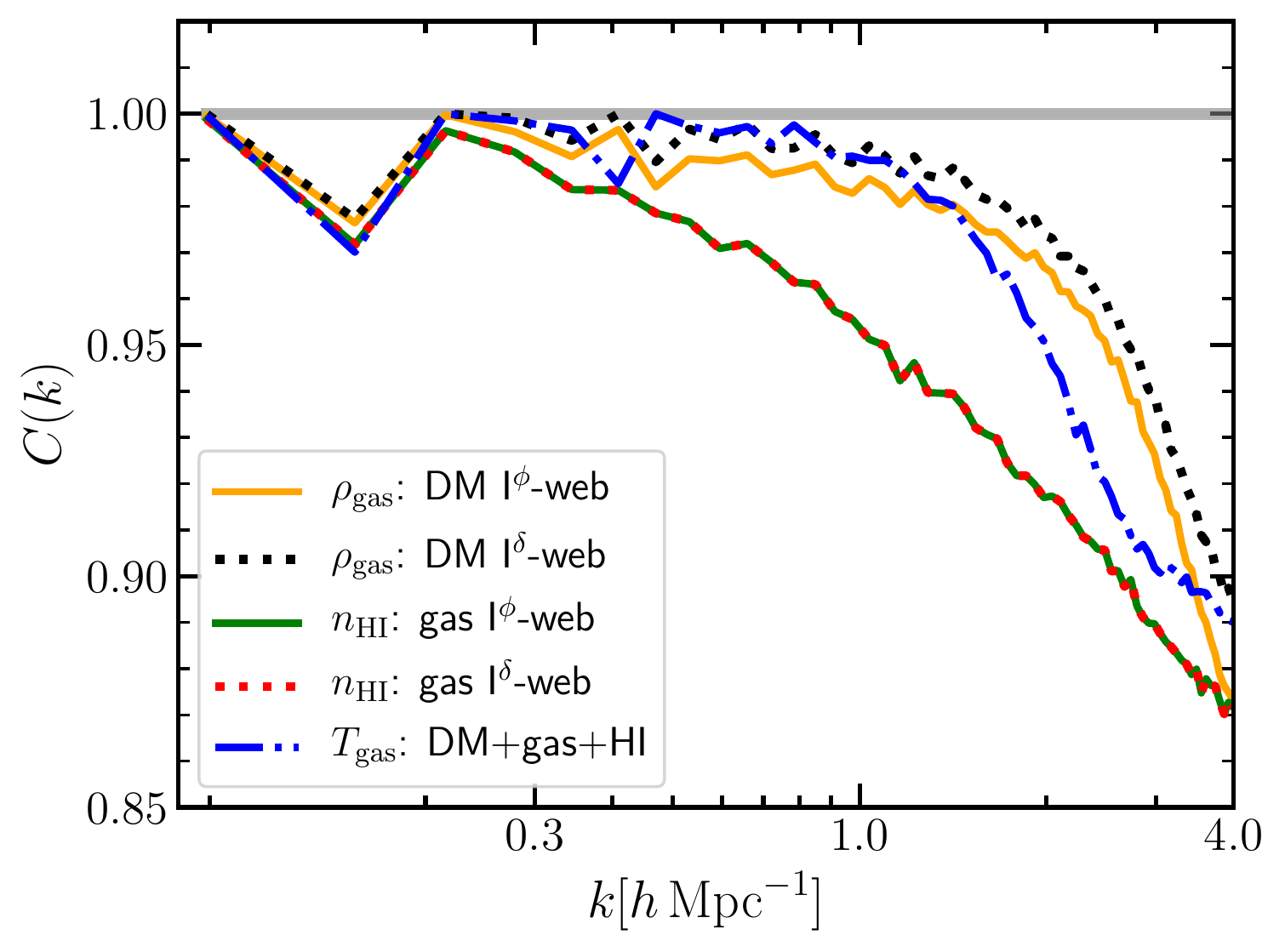}
\caption{\small{Correlation coefficients in Fourier space among the reference baryon fields and their reconstructions obtained by \texttt{BAM}, using different sets of underlying characterizing properties.}}
\label{fig:cross_corr_reconstr_refs}
\end{figure}
%%%%%%%%%%%%%%%%%%%%%%%%%%%%%%%%%%%%%%%%%%%%%%%%%%%%%%%

As previously pointed out, both the T-web and the I$^{\phi}$-web classification are examples of non-local properties of the dark matter field. Because of being defined through the gravitational potential, these characterizations correspond to long-range non-local properties. We can also explore short-range non-local terms \citep[see e.g.][]{McDonaldRoy2009,2020MNRAS.492.1614W}. The most simple example of such terms can be represented by the derivatives of the overdensity field, $\zeta_{ij} \equiv \partial_i\partial_j\delta$ \cite[see e.g.][]{Peacock1985,1986ApJ...304...15B}, whose principal and main invariants characterise the shape of local maxima (peaks) in a density field $\delta$. As for the I$^{\phi}$-web, we use the eigenvalues of the tensor $\zeta$ (labelled $I^{\delta}_{i}$, $i=1, 2, 3$ with $I^{\delta}_{1}=\nabla^{2}\delta$) to define a I$^{\delta}-$web classification.

In summary, we will assess the impact of different set of properties of the underlying DMDF in the \texttt{BAM} reconstruction process. These sets can be summarized as
\begin{itemize}
\item Local $\delta$: $\{\Theta\}=\{f(\delta) \}$
\item T-web : $\{\Theta\} =\{f(\delta),\, \rm{knots, filaments, sheets, voids} \}$
\item I$^{\phi}$-web : $\{\Theta\} =\lbrace f(\delta),\, g(I^{\phi}_2),\, g(I^{\phi}_3) \rbrace$
\item I$^{\delta}$-web : $\{\Theta\} = \lbrace f(\delta),\, g(I^{\delta}_1),\,g(I^{\delta}_2) \}$,
\end{itemize}
where the functions $f,g(x)$ represent non-linear transforms aiming at improving the mining of information in each variable $x$. We use $f(x)=\log(2+x)$ and $g(x)=2(x^{\alpha}-\gamma)/(\eta-\gamma)-1$ with $\gamma\equiv {\rm min}(x^{\alpha})$, $\eta \equiv {\rm max}(x^{\alpha})$ and $\alpha$ a free parameter, fixed in this work to $1/9$. The form of $f(x)$ has the usual form already used in BAM-I and BAM-II, whereas $g(x)$ aims at reducing the (large) range of values of the invariants $I^{\phi,\delta}_{i}$, by mapping them into the range $[-1,1]$, simplifying thus the binning strategy.
In practice we used $N_{\eta}=200$ and $N_{\{\Theta\}}=10^{6}$ bins for the total set of properties $\{\Theta\}$. We note that provided the total number of bins and the resolution of the mesh used to generate the different density fields explored in this work, it is reasonable to consider our learning procedure as an overfitting of the target density fields. Nevertheless, based on the findings of BAM-II we expect that the products of a calibration using one reference simulation can be used on independent density fields providing accurate ensemble statistics for the different baryon properties. This will be addressed in future publications.

%%%%%%%%%%%%%%%%%%%%%%%%%%%%%%%%%%%%%%%%%%%%%%%%%%%%%%%
\begin{figure*}
\center
\includegraphics[width=18cm]{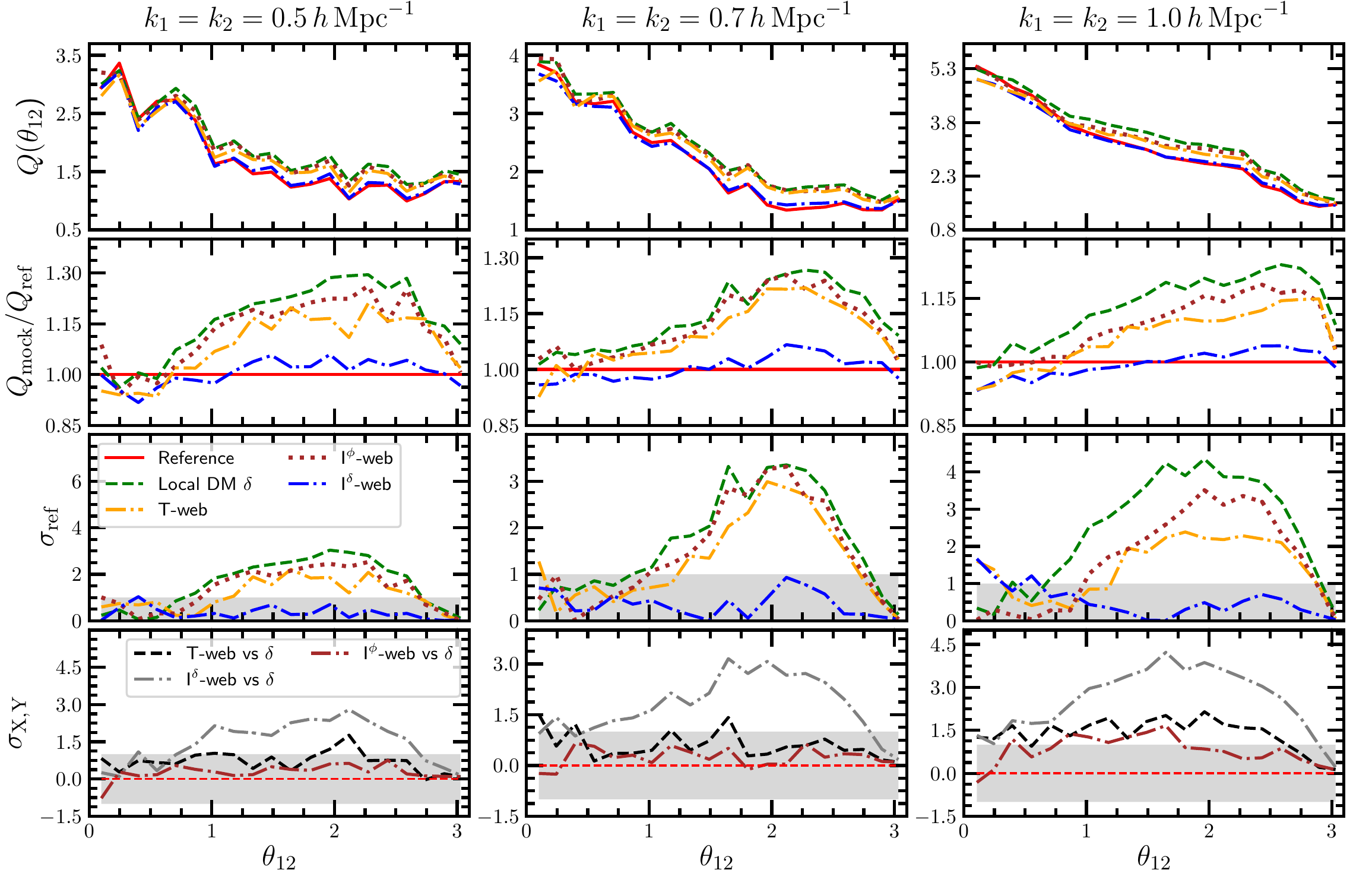}
\caption{\small{Three-point statistics in Fourier space of the distribution of ionized gas as reconstructed by \texttt{BAM}, using different characterization of the properties of the underlying dark matter field (see \S\ref{sec:bias}). The first row shows the reduced bi-spectrum $Q(\theta_{12})$ for isosceles configurations ($k_{1}=k_{2}$) probing different scales, as indicated. The second row shows the ratios between mock and reference bi-spectra $Q_{\rm mock}(\theta_{12})/Q_{\rm ref}(\theta_{12})$. The third row shows the statistical significance $\sigma_{\rm ref}$ of mock bi-spectra with respect to the reference bi-spectrum (see \S~\ref{sec:hydro_reproduction}). The bottom row shows the statistical significance $\sigma_{\rm X,Y}(\theta_{12})$ of the \texttt{BAM}
bi-spectrum obtained with bias model X with respect to the one generated with bias model Y, assuming that X is to be preferred over Y. Gray shaded areas represent the 1-$\sigma$ deviation.}} 
\label{fig:bk_gas_isosceles}
\end{figure*}
% %%%%%%%%%%%%%%%%%%%%%%%%%%%%%%%%%%%%%%%%%%%%%%%%%%%%%%

In Fig.~\ref{fig:slices_dm_prop} we show slices through the simulated volume with some of the properties of the DMDF aforementioned. In particular, it is evident how the information related to the tidal field encodes more information on the large-scale structure, while the invariant $I^{\delta}_{1}$ traces filamentary structure with an evident signature of local extrema. The capability of this invariant to resolve structures in the density field depends on the resolution of the mesh, in this case $\sim 0.8h^{-1}\rm{Mpc}$. Although a higher resolution can induce a more precise characterization of the maxima of the density field and its anisotropies, the information content can quickly saturate at the expense of a considerable increment in the computational time within the \texttt{BAM} reconstruction. Furthermore, lower mesh sizes smooth out the peaks, thus decreasing the amount of information in the invariants $I^{\delta}_{i}$. On the other hand, the second and third invariants, which correspond to terms of order $\mathcal{O}^{2}(I^{\delta})$ and $\mathcal{O}^{3}(I^{\delta}_{1})$ respectively, are less sensitive to the distribution of extreme values of the density field at this resolution, as can be seen from Fig.~\ref{fig:slices_dm_prop}. Indeed, our numerical experiments have shown that the set $\{I^{\delta}_{1},I^{\delta}_{2}\}$ is sufficient to achieve percent accuracy in the three-point statistics, and therefore we omit $I^{\delta}_{3}$ in the forthcoming analysis.

%%%%%%%%%%%%%%%%%%%%%%%%%%%%%%%%%%%%%%%%%%%%%%%%%%%%%%%
\begin{figure*}
\includegraphics[width=18cm]{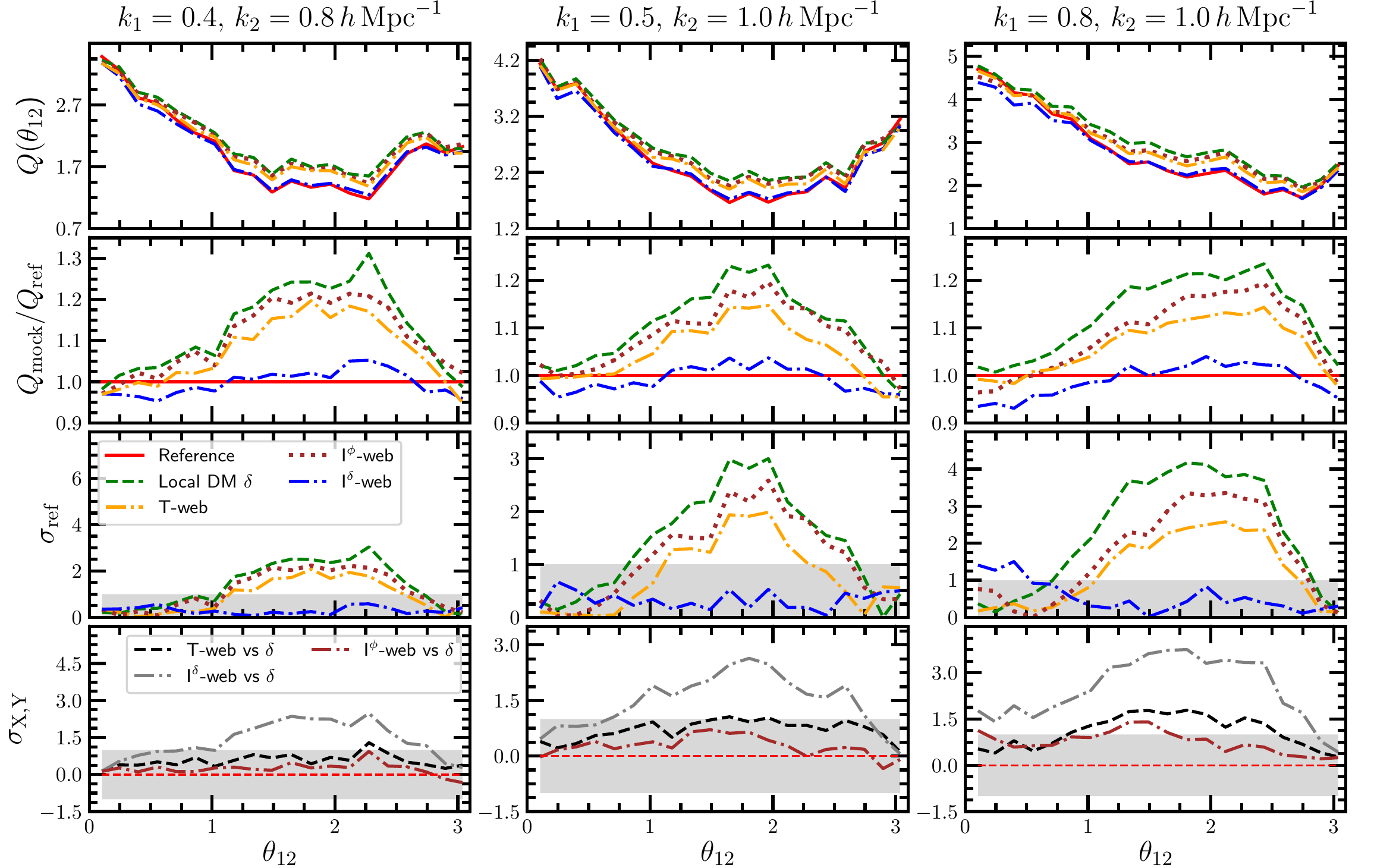}
\caption{\small{Reduced bi-spectrum from the distribution of ionized gas as reconstructed by \texttt{BAM}, using different characterization of the properties of the underlying dark matter field (see \S\ref{sec:bias}). This figure shows scalene configurations probing different scales, as indicated. See caption in Fig.~\ref{fig:bk_gas_isosceles} for further description.}} 
\label{fig:bk_gas_scalenes5}
\end{figure*}
%%%%%%%%%%%%%%%%%%%%%%%%%%%%%%%%%%%%%%%%%%%%%%%%%%%%%%%
%====================================================================================================
\subsection{Gas properties}

The first row of Fig.~\ref{fig:slices_ref_comp} shows examples of the spatial distribution of the gas properties through slices of the simulated volume. The slices show the distribution of fluctuations in the quantity $\log(2+\delta_{\eta})$, where $\delta_{\eta}=\eta/\langle \eta \rangle-1$ ($\eta=\rho_{\rm gas}, T_{\rm gas}, n_{HI}$). Visual inspection teaches us that each baryon property traces the filamentary structure designed by the underlying dark matter, although to a different degree. The most evident case is that of the ionized gas density $\rho_{\rm gas}$, whose distribution displays only slightly lower density contrasts, with a slightly more refined filamentary structure. The cosmic-web, as traced by the neutral hydrogen is less evident, blurred specially on the low density environments (i.e., sheets and voids), suppressing density contrast on filaments and highlighting only the highest peaks of the dark matter overdensity field. Finally, the gas temperature tends to break the filamentary structure not only by highlighting high density regions, but also increasing their apparent sizes.

The second and third rows in Fig.~\ref{fig:slices_ref_comp} show the distribution of DM and gas properties in different cosmic-web types (as defined in \S\ref{sec:bias}). While the gas density follows the same trend as the DM (albeit with broader distributions), the number density of HI and the gas temperature display multiple peaked distributions reflecting the presence of different thermodynamic phases of the gas. Such complex distributions, not being simple to disentangle analytically, are fully captured by the \texttt{BAM} machinery.

% ====================================================================================================
\subsection{Correlation between gas properties and dark matter properties}\label{sec:corr}
The complexity of the cosmological baryon phase-diagrams resolved by the hydro-simulation can be fully observed in Fig.~\ref{fig:hydro_cw}, where we present the joint probability distribution between pairs of gas properties and the dark matter density, again, separated in different cosmic-web types. The distinct behavior of the baryonic properties in these different environments \citep[see e.g.][]{2019MNRAS.486.3766M,2020arXiv201015139G} represents a physical motivation to perform a X-web type of decomposition, since it can highlight the environments populated by gas in its different phases. Notice for instance the well defined gas equation of state (gas density vs. gas temperature, right-most column in Fig.~\ref{fig:hydro_cw}) in voids (referred as "cool" gas ) in contrast to the behavior towards higher densities, where both "warm" and "hot" phases are present \citep[see e.g.][]{1997MNRAS.292...27H, 2001ApJ...552..473D, 2002A&A...388..741V}, associated with the gas shock-heated by gravitational collapse in knots and filaments \citep[see e.g.][for a recent detection of X-ray emission in filaments]{2020A&A...643L...2T}.

%%%%%%%%%%%%%%%%%%%%%%%%%%%%%%%%%%%%%%%%%%%%%%%%%%%%%%%
\begin{figure*}
\includegraphics[width=18cm]{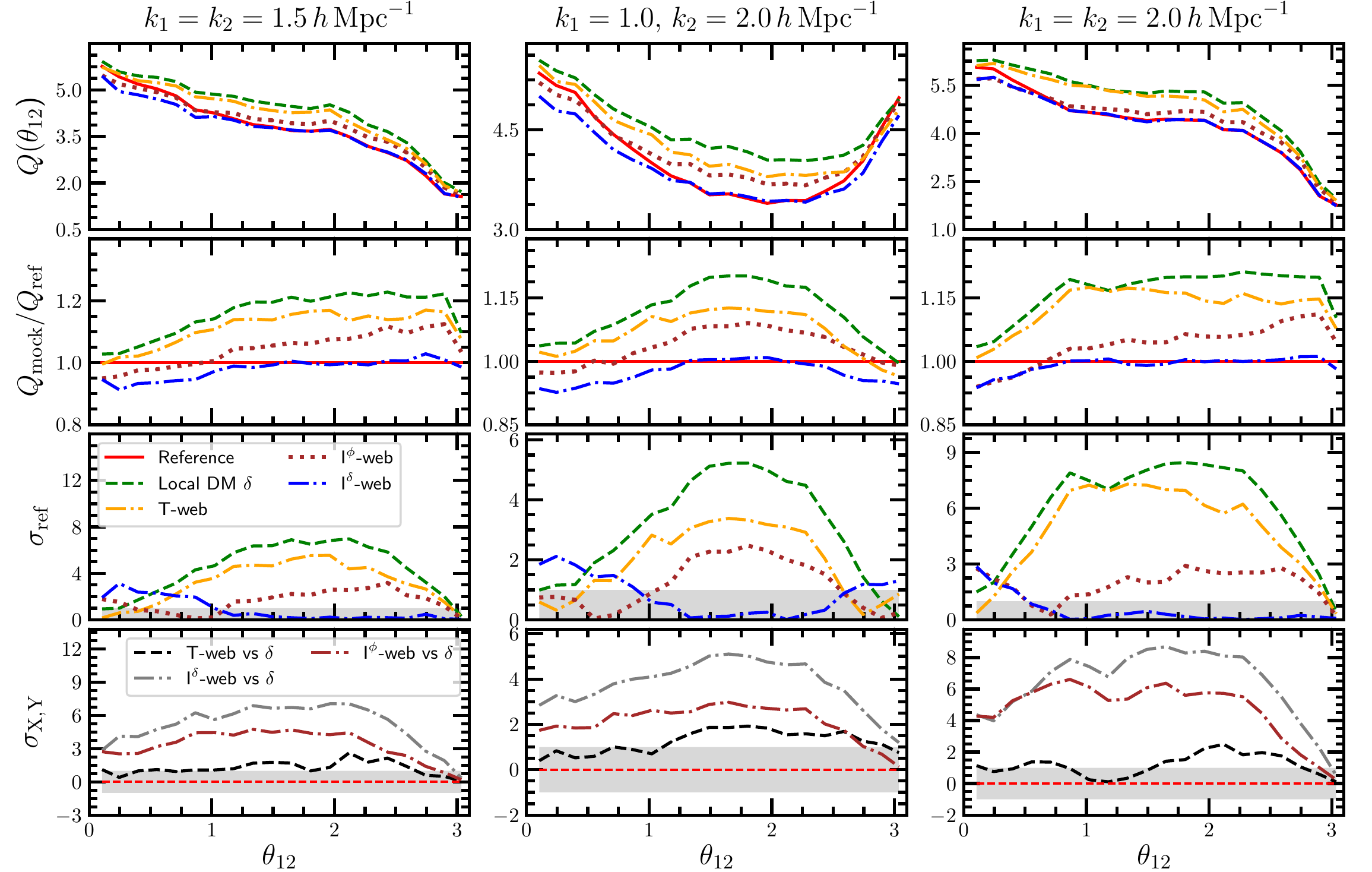}
\caption{\small{Reduced bi-spectrum from the distribution of ionized gas as reconstructed by \texttt{BAM}, using different characterization of the properties of the underlying dark matter field (see \S\ref{sec:bias}) and triangle configurations as indicated. See caption in Fig.~\ref{fig:bk_gas_isosceles} for further description.}} 
\label{fig:bk_gas_scalenes6}
\end{figure*}
%%%%%%%%%%%%%%%%%%%%%%%%%%%%%%%%%%%%%%%%%%%%%%%%%%%%%%%

The degree of correlation of the different gas properties among them and with respect to the underlying DMDF can be quantified through the correlation coefficient in Fourier space, shown in Fig.~\ref{fig:cross_corr_refs} \footnote{$C(k)=P_{\rm{XY}}(k)/\sqrt{P_{\rm{X}}(k)\,P_{\rm{Y}}(k)}$, where $P_{\rm{XY}}(k)\equiv \langle \delta_{X}(\vec{k})\delta^{*}_{Y}(\vec{k})\rangle $ is the cross-power-spectrum  of fields X and Y, while $P_{\rm{X}}(k)\equiv \langle|\delta_{X}(\vec{k})|^{2} \rangle $ denotes the power spectra of the $X$ field.}.
Panel~(a) in that figure shows the cross correlation between the three gas properties under scrutiny. While gas density and number counts of HI display a monotonically decreasing correlation towards small scales, the gas temperature correlates with the other properties in a more complex form, reflecting the different gas phases.

Panels (b) to (d) of Fig.~\ref{fig:cross_corr_refs} show the correlation of each of the gas properties with the properties of the underlying DMDF.
As expected from physical grounds and as anticipated by Fig.~\ref{fig:slices_ref_comp}, the gas density tightly correlates with the dark matter density. The tight correlation between gas and $I^{\delta}_{1}$ follows from the aforementioned correlation, and the fact that the power-spectrum  of the $I^{\delta}_{1}$ invariant is $\propto k^{4}P_{\rm dm}(k)$. The correlation of the gas-density with the invariants $I^{\phi}_{2,3}$ is weaker, approximately constant, though non-zero, while a strong anti-correlation is observed with the invariant $I^{\delta}_{2}$. The correlation between the gas density and the invariant $I^{\delta}_{3}$ follows approximately that of the invariant $I^{\phi}_{3}$.

The correlations of the number density of neutral hydrogen with the DM properties follows the same trend as the gas density, although with a mild tilt in the correlation with the invariants $I^{\phi,\delta}_{2,3}$ towards lower correlation on small scales. This can be interpreted as a signature of the biased HI distribution on small scales.

Finally, the correlation with the gas temperature displays, from a global view, the same behavior depicted before for distribution of gas and HI, although with some particular features. While on the largest scales the correlation with the dark matter is close to unity, on the range $0.2< k/(h\,{\rm Mpc}^{-1}) < 1$ it is instead dominated ($C(k)\sim 0.8$) by the invariants $I^{\phi}_{2,3}$ and $I^{\delta}_{2,3}$ (with $C(k)<0$ for $I^{\delta}_{3}$). This implies that the temperature distribution is more sensitive, on these scales, to the cosmic web (through $I^{\phi}_{i}$) and the shapes of local maxima (through $I^{\delta}_{i}$) \footnote{Notice from the bottom rows of Fig.~\ref{fig:slices_ref_comp} that the temperature distribution in knots and filaments differ only on high ($T>10^{5}$ K) temperatures.} This wave-number range can be roughly associated to comoving scales between $\sim 30$ and $\sim 5 h^{-1}$\,Mpc, corresponding to the typical separations within filaments and sheets. Note that at $k\sim k_{\rm nl}$, the correlation of the temperature with the dark matter starts to dominate. Provided the interpretation of the scale $k_{\rm nl}$, we can infer that such decline in the cross-correlation marks the transition from low-to high density regions in which cosmic-web related anisotropies are less strong.

% ====================================================================================================
% ====================================================================================================

\section{Reconstructing hydro-simulations} \label{sec:hydro_reproduction}
The current stage of the BAM machinery allows for the reconstruction of one tracer-property given the underlying DMDF. Extensions to parallel calibrations of two or more properties can be introduced at the expense of increasing the computational time (Balaguera-Antol\'{\i}nez $\&$ Kitaura, in preparation). Hence, to reconstruct a number of gas properties, we can establish a hierarchical approach, in which a main gas property is reconstructed first, based on the degree of correlation with the underlying DMDF. In view of the results presented in the previous section, we adopt the gas density as the main property to start the reconstruction with. Note that for the current research, the selection of a main property means that we will explore the all possible characterizations of the set $\{\Theta\}$ from the DMDF in order to reproduce the distribution of that \emph{main property}. In future works, aiming at producing mock catalogs, this will also mean that we can rely the reconstruction of other (less correlated with the DMDF) quantities on the reconstruction of the \emph{main property}.

%%%%%%%%%%%%%%%%%%%%%%%%%%%%%%%%%%%%%%%%%%%%%%%%%%%%%%%
\begin{figure*}
\hspace{-1cm}
\includegraphics[width=20cm]{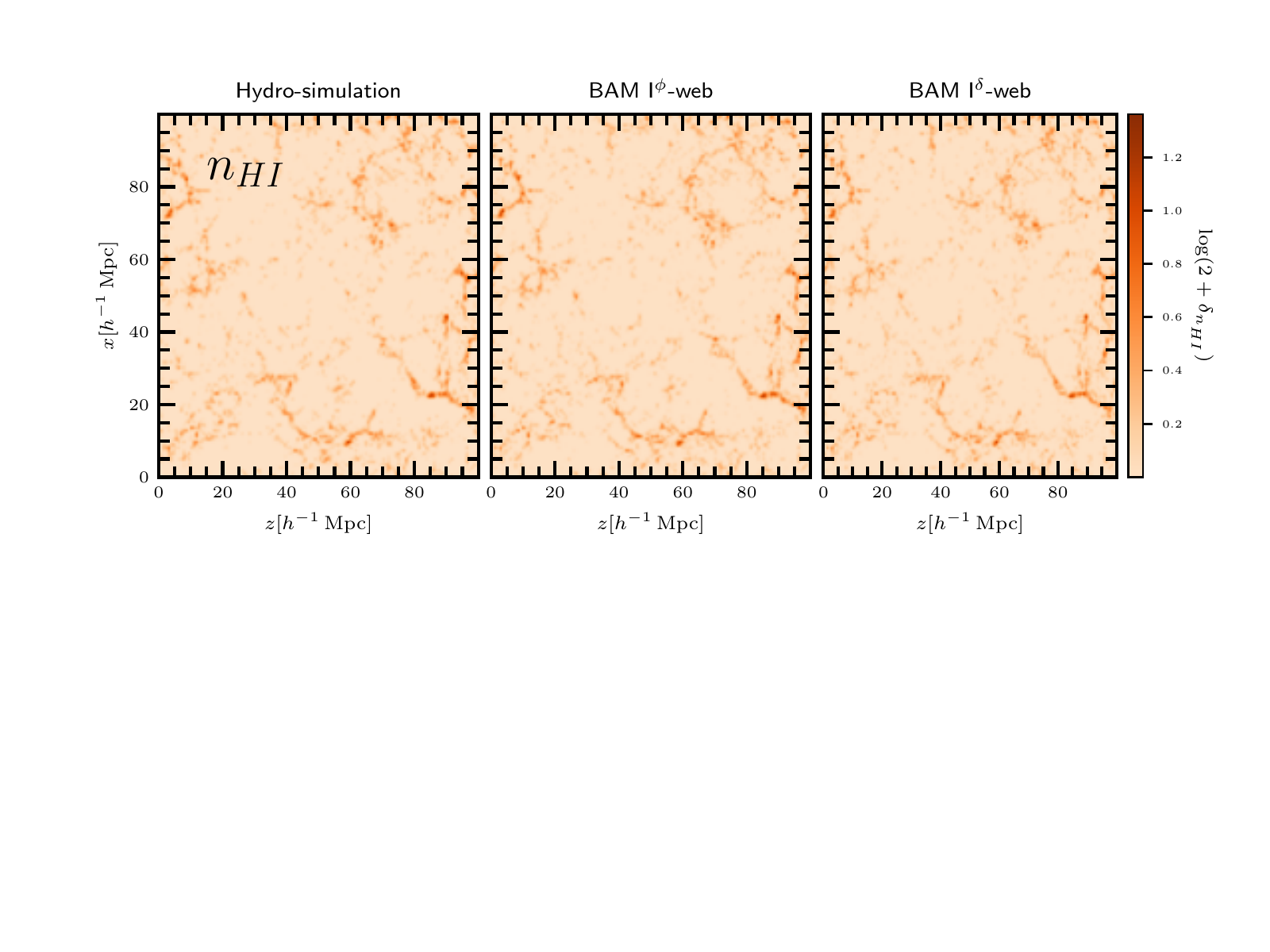}
\vspace{-7cm}
\caption{\small{Slices of $\sim 10h^{-1}$\,Mpc thick of neutral hydrogen density obtained from the reference simulation (top-left panel) and the \texttt{BAM} calibration procedure using different set of properties $\Theta$ evaluated over the gas distribution.}} 
\label{fig:slices_nH}
\end{figure*}
%%%%%%%%%%%%%%%%%%%%%%%%%%%%%%%%%%%%%%%%%%%%%%%%%%%%%%%

The reconstruction approach can be summarized as follows: 
\begin{itemize}
    \item \emph{Ionized gas density}: we determine the bias relation $\mathcal{P}(\delta_{\rm gas}| \mathcal{K}_{\rm gas}\otimes \{\Theta\}_{\rm dm})$ and the kernel $\mathcal{K}_{\rm gas}$, following the iterative procedure exposed in \S\ref{sec:bam}. Sampling this bias on the ($\mathcal{K}$-convolved) DMDF generates a new gas density field $\tilde{\delta}_{\rm gas}(\vec{r})$ following the procedure depicted by Eq.~(\ref{eq:sam_mock}). The iterative procedure ensures that the power-spectrum  of the field $\tilde{\delta}_{\rm gas}(\vec{r})$ follows, to percent precision, that of the field $\delta_{\rm gas}(\vec{r})$.
    
    \item \emph{Neutral hydrogen}: we assess the bias relation $\mathcal{P}(\delta_{\rm HI}| \delta_{\rm gas}, \mathcal{K}_{\rm HI}\otimes \{\Theta\}_{\rm dm})$ and the kernel $\mathcal{K}_{\rm HI}$. We then sample the DMDF to generate a new HI density field $\tilde{\delta}_{\rm HI}(\vec{r})$. Note however that given the tight correlation between the DM and the gas density, we can directly assess the scaling relation $\mathcal{P}(\delta_{\rm HI}| \delta_{\rm gas})$. In order to apply this to the iterative procedure, we determine the list of properties $\{\Theta\}_{\rm gas}$ of the gas density to obtain $\mathcal{P}(\delta_{\rm HI}| \mathcal{\tilde{K}}_{\rm HI}\otimes \{\Theta\}_{\rm gas})$ and the kernel $\mathcal{\tilde{K}}_{\rm HI}$ . Sampling the gas density field $\mathcal{\tilde{K}}_{\rm HI}\otimes \delta_{\rm gas}$ using this bias generates a new HI number density field $\tilde{\delta}_{\rm HI}(\vec{r})$. Here $\{\Theta\}_{\rm gas}$ denotes the set of properties listed in \S\ref{sec:bias}, this time computed based on the ionized gas density.

    \item \emph{Gas temperature}: we use the information of the gas density and the ionized hydrogen to reconstruct the information of the gas temperature, by computing the bias relation  $\mathcal{P}(T_{\rm gas}|\delta_{\rm HI},\delta_{\rm gas}, \mathcal{K}_{\rm T}\otimes \{\Theta\}_{\rm dm})$ and the kernel $\mathcal{K}_{T}$. Sampling the DMDF $\mathcal{\tilde{K}}_{\rm T}\otimes \delta_{\rm dm}$ with the resulting bias leads to a temperature density field $\tilde{\delta}_{T}(\vec{r})$. For this case, we will only use the local bias model $\Theta_{\rm dm}=\delta_{\rm dm}$. Note that the reason behind using $\rho_{\rm gas}$ and $n_{HI}$ to characterize the temperature bias is that we aim not only at reproducing the statistics of the spatial distribution, but also to keep track of thermodynamic relations between the baryon properties.
\end{itemize}

As commented in \S\ref{sec:bam}, a number of $\sim 200$ iterations is requested in order to achieve percent precision in the power-spectrum  (up the the Nyquist frequency) of each reconstruction. For the current mesh resolution ($128^{3}$), the CPU time requested for each iteration ranges from $\sim 2$ to $\sim 22$ seconds (on a 16- to 4-cores desktop/laptop) for local bias and X-web models respectively. Note that at the time of producing mock catalogs (i.e., after the calibration stage) using e.g. \texttt{ALPT} as gravity solver, \texttt{BAM} can replicate within few minutes the results of a $\sim 10^{5}$ CPU hours simulation. 
If we consider that we aim at producing of the order of ten snapshots we obtain a computing gain of  5 to 6 orders of magnitude, since  \texttt{BAM} requires $0.48\sim2\times10\times4\times22/3600$ CPU hrs (where we have assumed that the dark matter field estimation on a coarse mesh introduces a factor 2 more computing time and 4-cores are used) vs $1.04\times 10^{5}$ CPU hrs in the case of the hydrodynamic simulation.

%%%%%%%%%%%%%%%%%%%%%%%%%%%%%%%%%%%%%%%%%%%%%%%%%%%%%%%
\begin{figure*}
\includegraphics[width=18cm]{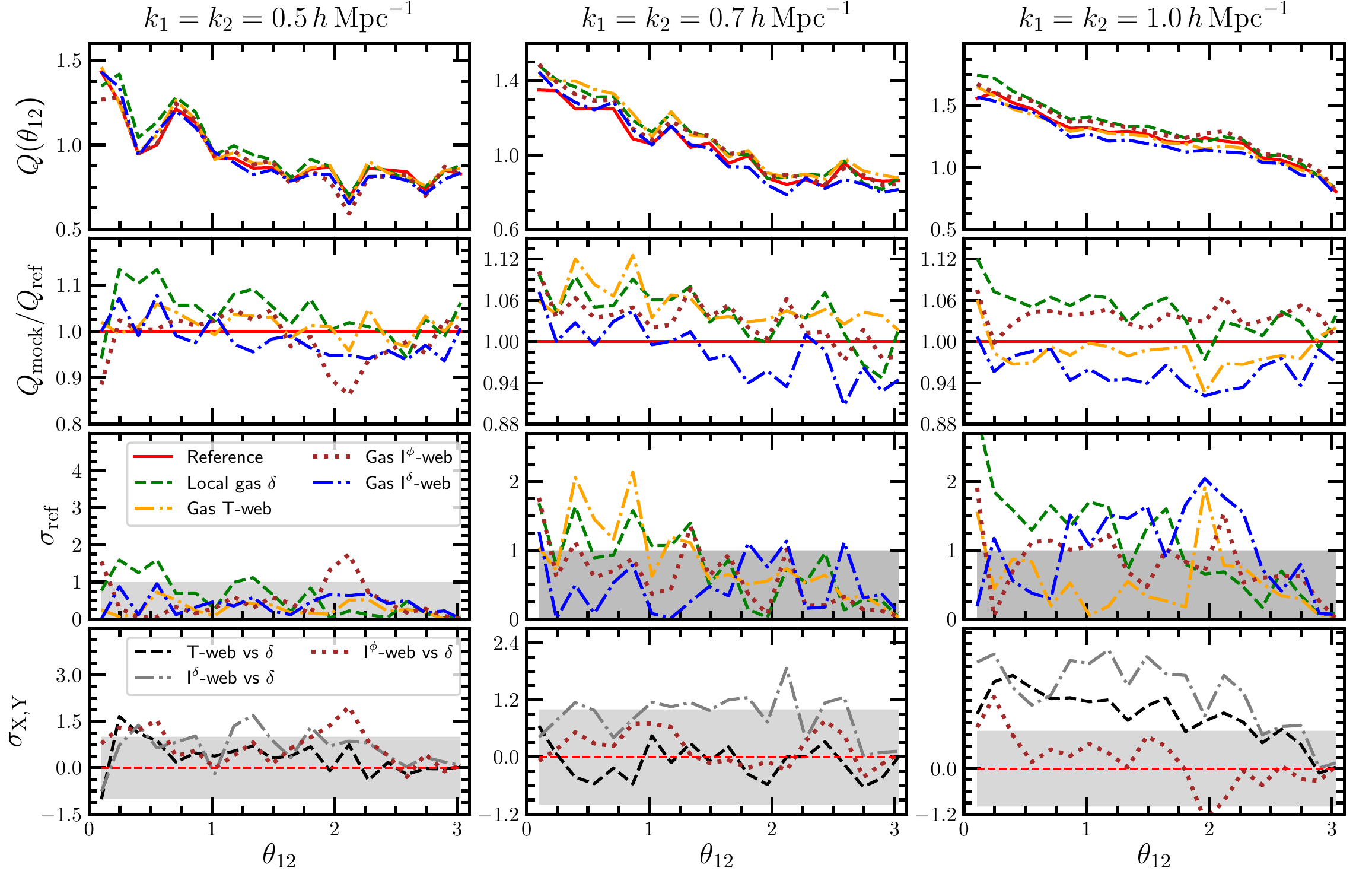}
\caption{\small{Three-point statistics from the distribution of HI as reconstructed by \texttt{BAM}, using different characterization of the properties of the gas distribution (see \S\ref{sec:bias}). The first row shows the reduced bi-spectrum $Q(\theta_{12})$ for isosceles configurations $k_{1}=k_{2}$ probing different scales as indicated. The second row shows the ratios between mock and reference bi-spectra $Q_{\rm mock}(\theta_{12})/Q_{\rm ref}(\theta_{12})$. The third row shows the statistical significance $\sigma_{\rm ref}$ of mock bi-spectra with respect to the reference bi-spectrum (see \S\ref{sec:hydro_reproduction}). The bottom row shows the statistical significance $\sigma_{\rm X,Y}(\theta_{12})$ of the \texttt{BAM} bi-spectrum obtained with bias model X with respect to the one generated with bias model Y, assuming that X is to be preferred on Y. Gray shaded areas represent the $1-\sigma$ deviation.}}
\label{fig:bk_nHI_isosceles}
\end{figure*}
%%%%%%%%%%%%%%%%%%%%%%%%%%%%%%%%%%%%%%%%%%%%%%%%%%%%%%%

We now present the results from each of these reconstruction. As previously mentioned, since the power-spectrum  is the target statistics within the \texttt{BAM} procedure, it's resulting signal is expected to follow to percent accuracy that of the reference regardless, in principle, the set of properties $\{\Theta\}$. Discriminating among the different options for such characterization at the two-point statistics level can be done through the shape of the resulting kernel $\mathcal{K}$, although no quantitative assessment of the information encoded in each choice of $\{\Theta\}$ is possible in that scenario\footnote{The kernel can provide information on the large-scale bias associated to properties not included in the multi-dimensional bias description characterized by the set $\{\Theta\}$}. Instead, the three-point statistics, not being part of the \texttt{BAM} calibration, is highly sensitive to the choice of the set $\{\Theta\}$. We therefore assess the information content in the different characterizations $\{\Theta\}$ through the signal of the reduced bi-spectrum\footnote{We use the code \url{https://github.com/cheng-zhao/bispec} made available by Cheng Zhao.} of the \texttt{BAM} reconstruction compared to the reference signal. The reduced bi-spectrum is defined as $Q(\theta_{12})=B(k_1,k_2,\theta_{12})/P(k_{1})P(k_{2})P(k_{3})$, where $\theta_{12}= {\rm arccos}(\vec{k}_{1}\cdot \vec{k}_{2})$ and $\vec{k}_{1}+\vec{k}_{2}+\vec{k}_{3}=0$. We explore isosceles and scalene triangle  configurations probing different scales within the simulated volume, with $k_{1}=k_{2}=(0.5, 0.6, 0.7 )\,h{\rm Mpc}^{-1}$ and $(k_{1},k_{2})=(0.4,0.8), (0.5,1.0), (0.8, 1.0)\,h{\rm Mpc}^{-1}$. These scales are sensitive to BAO and redshift space distortions \cite[see e.g.][]{2008PhRvL.100i1303C,2012MNRAS.422..926M,2020arXiv201103558R}. For the reconstruction of the ionized gas, we explore smaller scales, up to $k=2\,h\,$Mpc$^{-1}$.

%%%%%%%%%%%%%%%%%%%%%%%%%%%%%%%%%%%%%%%%%%%%%%%%%%%%%%%
\begin{figure*}
\includegraphics[width=18cm]{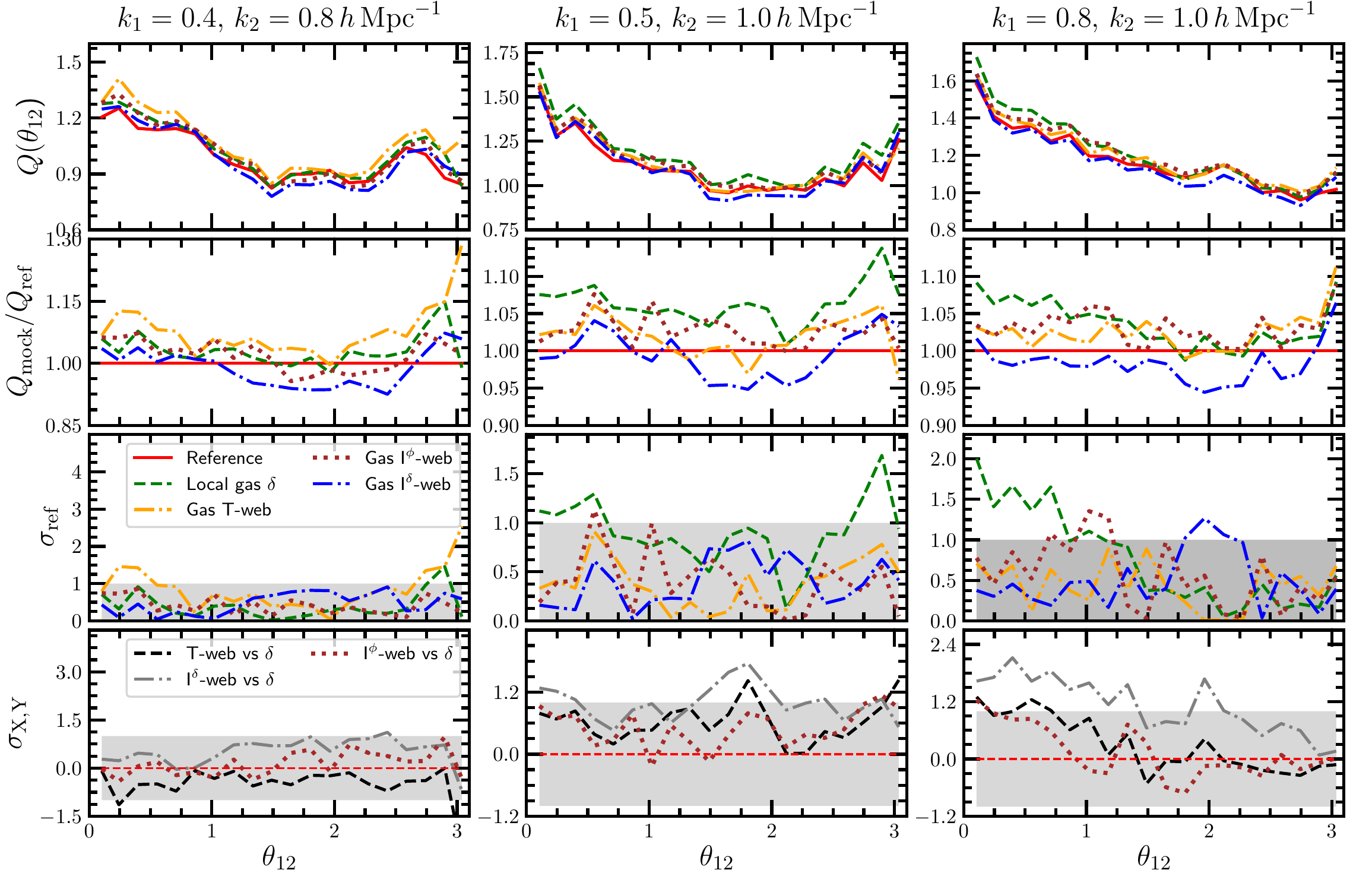}
\caption{\small{Reduced bi-spectrum from the distribution of HI reconstructed by \texttt{BAM}, using different characterization of the properties of the underlying dark matter field (see \S\ref{sec:bias}). This figure shows scalene configurations probing different scales, as indicated. See Fig.~\ref{fig:bk_nHI_isosceles} for further descriptions.}} 
\label{fig:bk_nHI_scalenes}
\end{figure*}
%%%%%%%%%%%%%%%%%%%%%%%%%%%%%%%%%%%%%%%%%%%%%%%%%%%%%%%

The uncertainties in the bi-spectrum are estimated from the approximation $\sigma^2_{\rm{B}} \approx (s_{\rm{B}}/V_{\rm B})P(k_1)P(k_2)P(k_3)$, \citep[e.g.][]{fry_article,Scoccimarro:2000sn}, where $s_{\rm{B}}=2,1$ for isosceles and scalene configurations respectively, and $V_{\rm{B}}\approx 8\pi^2 k_1 k_2 k_3$. The uncertainty $\sigma_{\rm{Q}}$ on reduced bi-spectrum is then obtained by propagating these errors under the assumption that the errors in the power-spectrum  can be derived from a Gaussian approximation. For each reconstructed property, we will quantify the statistical significance of the bi-spectrum signal between different sets of $\{\Theta\}$ (say, X and Y) through 
$\sigma_{\rm{X,Y}}(\theta_{12})\equiv \left(Q_{\rm{X}}(\theta_{12})-Q_{\rm{Y}}(\theta_{12})\right)/\sqrt{2}\sigma_{Q}$, while the statistical significance of each set $\{\Theta\}$ with respect to the reference will be quantified as $\sigma_{\rm{ref}}=|\sigma_{\rm{X,Y}}|$ where $Y$ stands for the reference signal.

% ====================================================================================================

\subsection{Ionized gas}
\label{sec:gas}
In Fig.~\ref{fig:slices_gas}  we show the slices of the simulated volume showing the distribution of ionized gas as obtained from the \texttt{BAM} calibration using different set of dark matter properties as listed in \S\ref{sec:bias}. All models used within the iterative procedure reconstruct the original cosmic-web structure as probed by the ionized gas, with minor small-scale differences, specially visible within the filamentary structures.
Panel (a) in Fig.~\ref{fig:pk_final} shows the corresponding power-spectrum  for the gas distribution obtained from the different reconstructions. The black-dashed line in that plot shows an example of the power-spectrum  obtained by sampling an estimate of the gas bias following Eq.~(\ref{eq:sam_raw})  (using the local-bias model), i.e prior to the iterative procedure. This evidences the need for the kernel in order to correct for the observed $\sim 15\%$ excess of power.

The second row-panels in that plot show the ratio $P_{\rm{mock}}(k)/P_{\rm{ref}}(k)$ between the \texttt{BAM} reconstruction and reference power-spectrum. The models implemented lead to gas-power spectra which display random fluctuations of the order of $\sim 1-2\%$ with respect to the reference, up to the Nyquist frequency. In all cases, the averaged residuals are well below $1\%$. Note that the \texttt{BAM} machinery has provided the same level of accuracy starting from different models. 
The difference in the implementation of the models is encoded, on one hand, in the bias and the kernel obtained after each iterative procedure, and in the other hand, in the signal of the three-point statistics.

The bottom panels in Fig.~\ref{fig:pk_final} show the resulting kernel obtained from the iterative process using different sets of $\{\Theta\}$. In general, the kernel of the gas reconstruction (as that for the HI) has a constant behavior on large scales ($\lim_{k\to 0}\mathcal{K}(k)=\mathcal{K}_{0}=$ constant, depending on the chosen description of the DM properties) and a scale-dependent shape ($\mathcal{K}(k)\to k^{\alpha}$, $\alpha>0$) on small scales ($k>k_{\rm nl}$). 
As anticipated in \S~\ref{sec:bam}, this wavenumber marks the scale at which the kernel accounts for missing local contributions to the bias (large scales). 
The missing information can be effectively replicated by decreasing the amplitude of the density fluctuations on large scales (by a factor $\mathcal{K}_{0}$) while modulating these towards a more populated high-$\delta_{\rm dm}$ probability distribution. In particular, for the case of the gas density, the calibration with the I$^{\delta}$-web model generates a kernel with amplitude on large-scales closer to unity, suggesting that such set of properties encodes more information of the gas spatial distribution than its tested counterparts. This is indeed concluded from the three-point signal, as we discuss below.

%%%%%%%%%%%%%%%%%%%%%%%%%%%%%%%%%%%%%%%%%%%%%%%%%%%%%%%
\begin{figure*}
    \hspace{-2cm}
    \includegraphics[width=21cm]{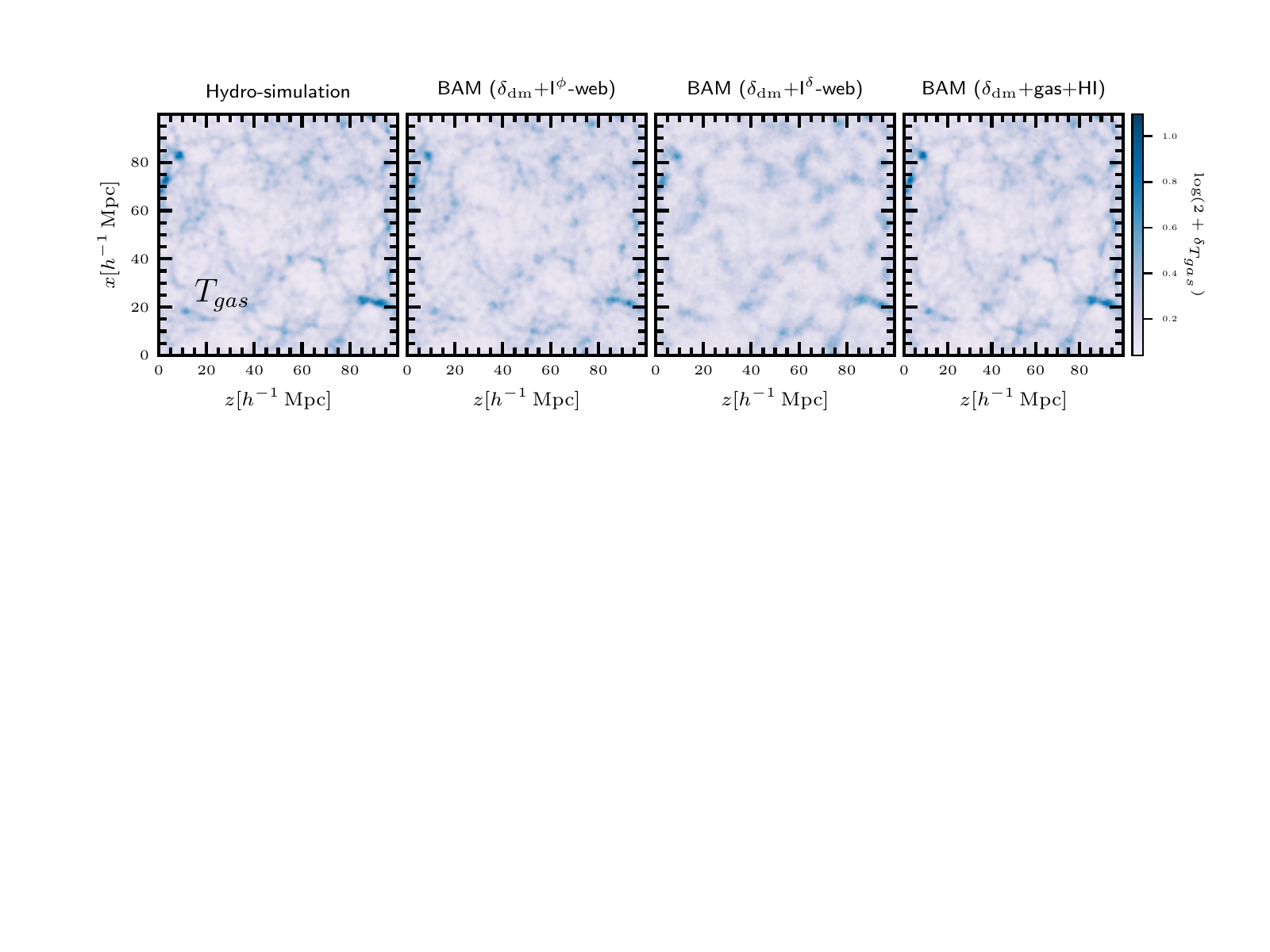}
\vspace{-9cm}
    \caption{\small{Slices of $\sim 10h^{-1}$\,Mpc thick of ionized gas temperature obtained from the reference simulation (left panel) and the \texttt{BAM} calibration procedure using different characterizations $\{\Theta\}_{\rm dm}$ such as the I$^{\phi,\delta}$-web and the information from the gas density and HI number density.}}
    \label{fig:temperature_slices}
\end{figure*}
%%%%%%%%%%%%%%%%%%%%%%%%%%%%%%%%%%%%%%%%%%%%%%%%%%%%%%%

Alternative qualitative assessments of the \texttt{BAM} reconstruction can be made through the cross-correlation between the reconstructed fields and the original ones. This is shown in Fig.~\ref{fig:cross_corr_reconstr_refs}, where we denote with solid (orange) and dotted (black) lines the correlation coefficients among the reference density field and its reconstruction for the gas, using the I$^{\phi}$-web and I$^{\delta}$-web models respectively. In both cases $C(k)>0.95$ at $k\lesssim 2.0\,h \rm{Mpc}^{-1}$, with  $C(k)\sim 0.87$ at the Nyquist frequency. The I$^{\delta}$-web characterization of the DMDF properties yields higher correlations at all wavenumbers $k$.

Figures \ref{fig:bk_gas_isosceles}, \ref{fig:bk_gas_scalenes5} and \ref{fig:bk_gas_scalenes6} show the measurements of the reduced bi-spectrum from the \texttt{BAM} calibrations, using each of the sets of DM properties $\{\Theta\}$ listed in \S\ref{sec:bias}. In general, a maximum  deviation of $\sim 10\%$ with respect to the reference are observed. The results shown suggest that using only local information does not accurately reproduce the reference three-point statistics at all the considered scales. Furthermore, and contrary to the reconstruction of the distribution of dark matter halos \citep[see e.g.][]{Kitaura2020}, the I$^\phi$-web classification leads to a mild improvement in the accuracy, while the invariants $I^{\delta}_{1,2}$ provide a better signal of reduced bi-spectrum from the reconstructed gas distribution (as suggested by the kernel in Fig.~\ref{fig:pk_final}). For this choice of $\{\Theta\}$, we find a difference between the reference and reconstructed bi-spectra below the (approximated) error bars on scales $k\lesssim 1.0\,h\rm{Mpc}^{-1}$, as can be inferred from the third row in Figs.~\ref{fig:bk_gas_isosceles} and \ref{fig:bk_gas_scalenes5}. 
In terms of the explored sets of $\{\Theta\}$, we find a $4-5\sigma$  significance for the I$^\delta$-web model with respect to the local $\delta$ bias model. This finding can be interpreted as a smoking gun for the need of short-range biasing modeling to account for the clustering of baryons at the scales probed by the hydro-dynamical simulation.

Interestingly, on smaller scales, as shown in Fig.~\ref{fig:bk_gas_scalenes6}, the best reconstructions are not only accounted by the $I^{\delta}-$web. Instead, the $I^{\phi}-$web can account for the reference signal on angles $\theta_{12}$ on which $I^{\delta}-$web considerably deviates from the reference. This would indicate that a full treatment involving both $I^{\delta}$ and $I^{\phi}-$web descriptions can reconstruct, to high accuracy, the three-points statistics of the gas distribution on such small scales. We leave this research for future publications.

% ====================================================================================================
\subsection{Neutral Hydrogen}
\label{sec:nHI}

As pointed out in \S\ref{sec:bam}, the tight correlation between the gas and the dark matter distribution (see e.g. Fig.~\ref{fig:hydro_cw}) motivates us to reconstruct the distribution of HI from the gas density, instead of the dark matter density. With this approach, we aim at encapsulating underlying statistical information of the physics related to the gas and HI distribution not present in the correlation between DM and gas. Accordingly,  we evaluate the set of properties $\Theta$ based on the gas density $\rho_{\rm gas}$ and its corresponding overdensity. 

%%%%%%%%%%%%%%%%%%%%%%%%%%%%%%%%%%%%%%%%%%%%%%%%%%%%%%%
\begin{figure*}
\centering
\includegraphics[width=18cm]{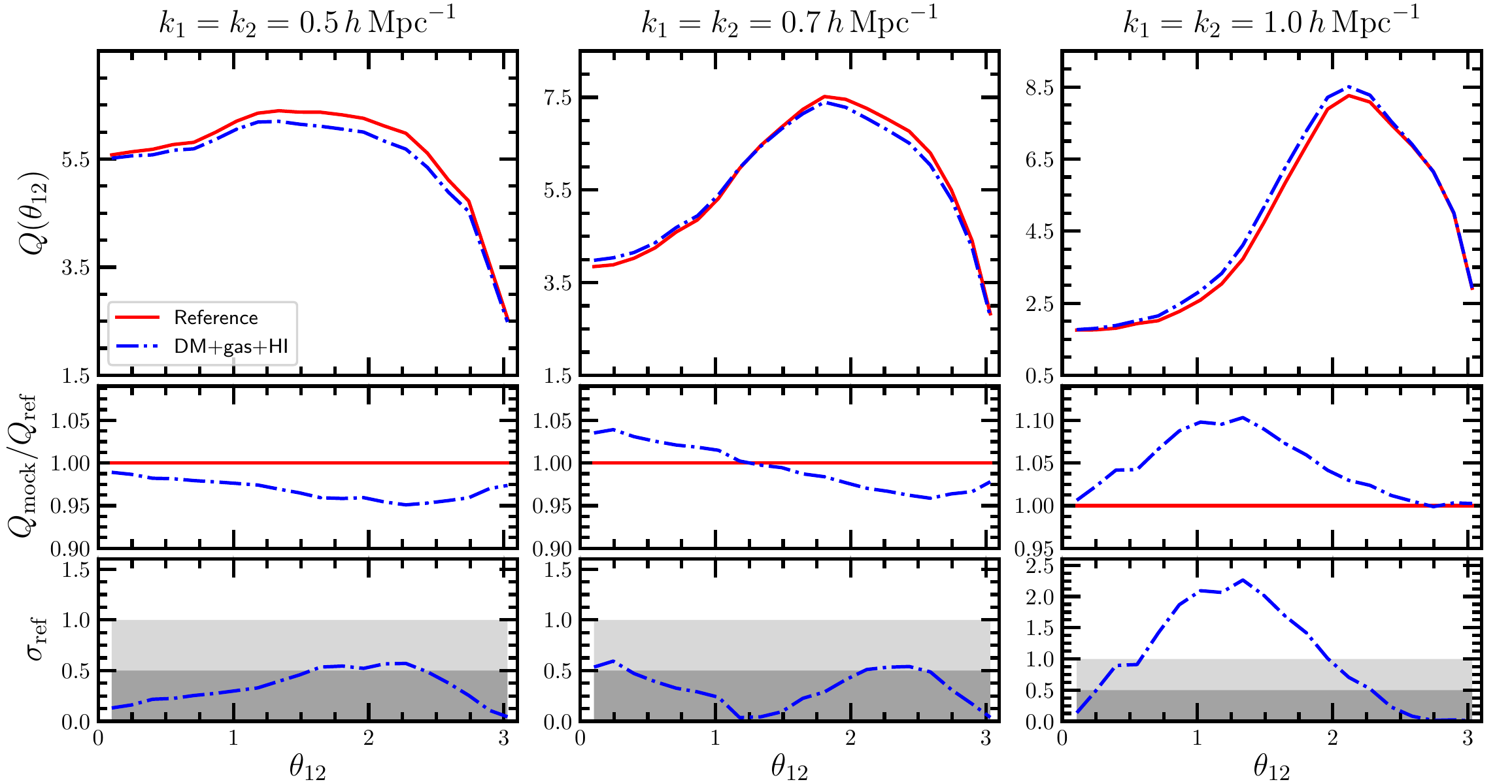}
\caption{\small{Three-point statistics from the distribution of temperature as reconstructed by \texttt{BAM}, using the local properties of the underlying dark matter, ionized gas and neutral hydrogen fields (see \S\ref{sec:hydro_reproduction}). The first row shows the reduced bi-spectrum $Q(\theta_{12})$ for wave-numbers $k_{1}$ and $k_{2}$ fixed as indicated. The second row shows the ratios between mock and reference bi-spectra $Q_{\rm mock}(\theta_{12})/Q_{\rm ref}(\theta_{12})$. The bottom row shows the statistical significance $\sigma_{\rm ref}$ of mock bi-spectra with respect to the reference bi-spectrum (see \S\ref{sec:hydro_reproduction}). Gray shaded areas represent the $0.5\,\sigma$ and $1$-$\sigma$ deviations.}}
\label{fig:bk_temp_isosceles}
\end{figure*}
%%%%%%%%%%%%%%%%%%%%%%%%%%%%%%%%%%%%%%%%%%%%%%%%%%%%%%%

The panel (b) in Fig.~\ref{fig:pk_final} shows the power-spectrum  of the reconstructed HI field, displaying an average residuals of $\sim 0.8 \%$ in all variations of the set $\{\Theta\}$. Such agreement can be visually confirmed from Fig.~\ref{fig:slices_nH}, which shows slices of the simulated volume comparing the reconstruction, using the I$^\phi$-web and the I$^\delta$-web characterisations based on the DMDF. Compared to the power-spectrum  of the reconstruction of the gas density field, the corresponding statistics for the HI has slightly larger differences with respect to the reference (e.g. $\sim 3\%$ at $k\sim 0.3 h{\rm Mpc}^{-1}$ for HI, compared to $\sim 1.5\%$ of the gas at the same scale). The bottom plot of the panel (b) in Fig.~\ref{fig:pk_final} shows the kernel obtained with the different choices of DM properties. Contrary to the gas reconstruction, in this case the I$^{\delta}$-web characterization is less informative as is confirmed by the three-point signal. Also, the kernels for the HI reconstruction is slightly more noisy than that obtained for the gas density. This feature, together with the slightly larger discrepancy with respect to the reference power-spectrum  (compared to the gas reconstruction) can be understood as a consequence of the lower correlation between HI and the DMDF, as shown in Fig.\ref{fig:cross_corr_refs}.

In terms of the cross-correlation between the reconstructed fields and the original ones, the solid-green and dotted-red lines in Fig.~\ref{fig:cross_corr_reconstr_refs} show the correlation coefficients for the HI number density field, using the $I^{\phi}$-web and $I^{\delta}$-web set of properties of the underlying gas density field, respectively. In both cases $C(k)>0.90$ at $k\lesssim 2.0h\rm{Mpc}^{-1}$ and $C(k)>0.85$ up to the Nyquist frequency.

Figures~\ref{fig:bk_nHI_isosceles} and \ref{fig:bk_nHI_scalenes} show the measurements of reduced bi-spectrum in the different choices for the set $\{\Theta\}$ (based on the gas distribution), and probing different triangle configurations. We observe that the I$^\delta$-web decomposition generates in general a more accurate description of the three-point statistics of the HI distribution. The I$^{\phi}$-web approach, as well as the local bias dependency perform still (within a $1-3\sigma$ difference) with respect to the reference. We have verified that on triangle configurations probing smaller scales (i.e., $k_{1}=k_{2}=2 h\,{\rm Mpc}^{-1}$), the I$^\delta$-web model yields $< 1\sigma$ deviations with respect to the reference.

% ====================================================================================================
\subsection{Gas temperature}
\label{sec:temperature}

As anticipated at the beginning of this section, the \texttt{BAM} reconstruction of the gas temperature has been performed using the information of the local dark matter density, the gas density and the number density of HI. The rationale behind this choice is that we aim not only at reproducing the statistics of the spatial distribution, but also the thermodynamic relations between the baryon properties. %We leave a study including additional non-local bias relations (through the I$^{\delta}$-web and I$^{\phi}$-web) for future work, as these demand more computational memory resources\footnote{At this stage, all the \texttt{BAM} calculations shown in this work can be performed  on a desktop/laptop with 16 Gb RAM and 4 cores.}.

A visual inspection confirms very similar patterns comparing slices of the reference simulation to the reconstructed temperature field, as shown in  the rightmost panel of Fig.~\ref{fig:temperature_slices}. 
The same figure shows the \texttt{BAM} reconstruction using the I$^{\phi,\delta}$-web in the middle panels. These do not replicate the cosmic-web as accurately as the case in which baryon information is implemented.  We find that structures are fragmented in the I$^\phi$-web case, and smoothed out in the I$^\delta$-web case. In fact, they yield non-negligible deviations in the summary statistics (e.g. bias in the power-spectrum  of about $\gtrsim 20\%$). This implies that, even though the temperature distribution is sensitive to the non-local bias induced by the cosmic-web, the main source of information is encoded in the local representation of dark matter and baryon properties. Therefore, we will focus on the rightmost reconstruction case $\delta_{\rm dm}$+gas+HI. 

Panel (c) in Fig.~\ref{fig:pk_final} shows the resulting power-spectrum  of the temperature $T_{\rm gas}$ field for this latter case. After having reached a stationary state in the calibration procedure, the resulting residuals (of the order of $3\%$ up to the Nyquist frequency) generate a power-spectrum  which systematically deviates from that of the reference, being within about 2\% up to $k\sim 1\,h\,$Mpc$^{-1}$, with a maximum deviation of $\sim 6 \%$ at $k\sim 2.5\,h\,$Mpc$^{-1}$. The same panel shows how a temperature reconstruction using only the local DM density fails at reproducing the power-spectrum  at all scales (green-dashed line). Despite the systematic bias observed in the reconstruction, it is remarkable how the temperature field is taken from the first iteration (black-dashed line) displaying $\sim 1.5$ orders of magnitude difference with respect to the reference at $k=0.3\,h{\rm Mpc}^{-1}$, to the $\sim 2\%$ difference at the same scale.

The bottom plot of the corresponding panel shows the kernel for the calibration of the temperature field. Contrary to that observed for the gas density and neutral hydrogen, the kernel for the temperature displays values larger than unity. That is, the reconstruction of gas temperature based on dark matter (green-dashed line) and/or the set dark matter plus baryon properties (blue-dashed-dotted line) demands a $\sim 10^{2}$ increment in the amplitude of the DM density perturbations. Again, as pointed out before, such figures are a consequence of the complexity of the distribution of gas temperature and its low correlation with the underlying DMDF, as well as need for more physical information (beyond dark matter, gas density and neutral hydrogen), such as shock heating and feedback from supernovae and active galactic nuclei.

In terms of the correlation coefficients between the reference and the reconstructed fields, the blue dash-dotted line in Fig.~\ref{fig:cross_corr_reconstr_refs} shows how using local information on dark matter, gas ad HI, we obtain correlation coefficients of $C(k)>0.95$ at $k\lesssim 2.0h\rm{Mpc}^{-1}$ and $C(k)\sim 0.90$ at the Nyquist frequency.

\begin{figure*}
\includegraphics[width=18cm]{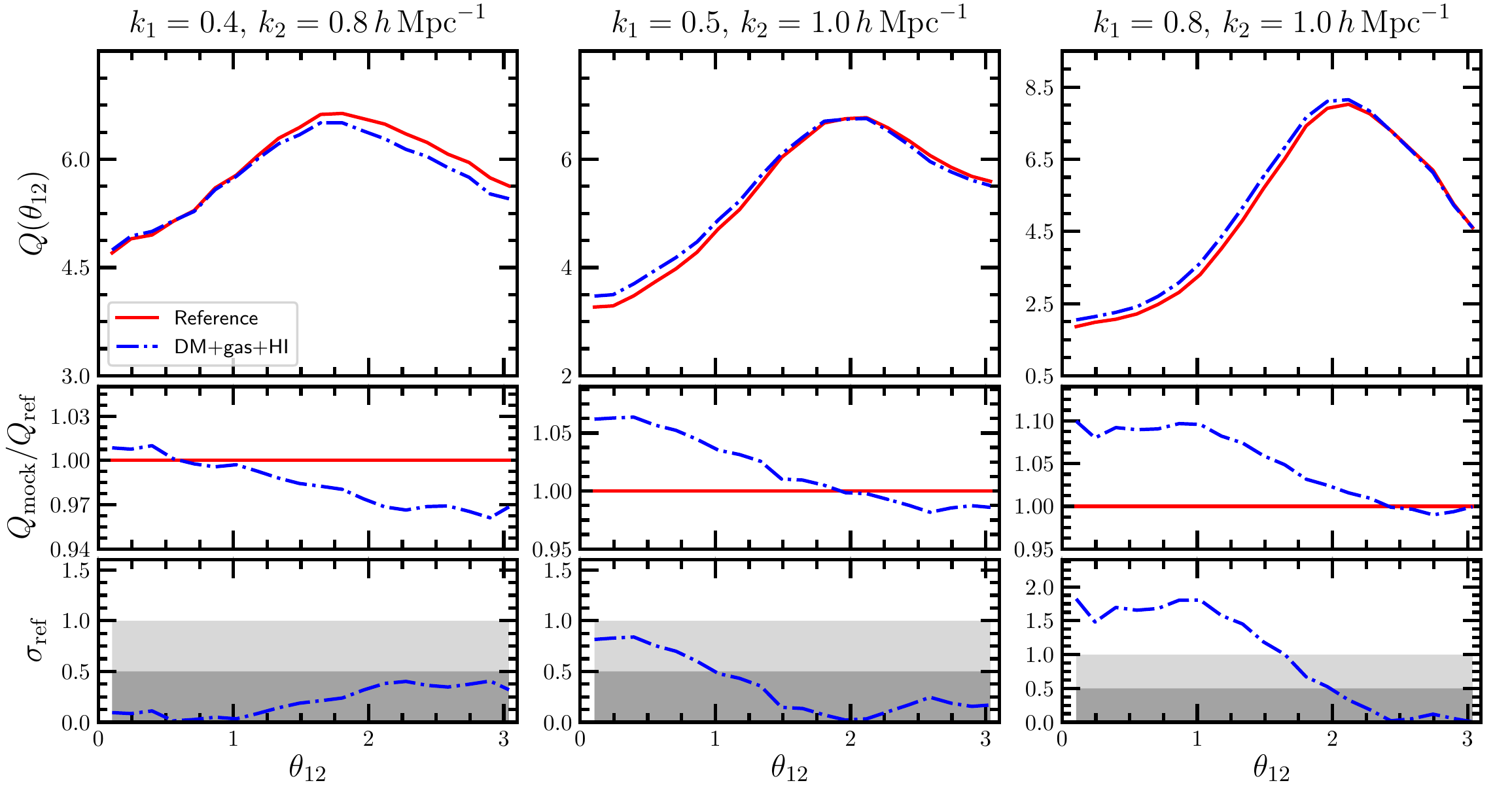}
\caption{\small{Reduced bi-spectrum from the distribution of gas temperature reconstructed by \texttt{BAM}. This figure shows scalene configurations probing different scales, as indicated. See Fig.~\ref{fig:bk_temp_isosceles} for further description.}} 
\label{fig:bk_temp_scalenes}
\end{figure*}

The corresponding reduced bi-spectrum resembles well the reference one, as shown in Figs.~\ref{fig:bk_temp_isosceles} and \ref{fig:bk_temp_scalenes}, where a $\sim 5\%-10\%$ differences with respect to the reference are observed, in the different probed configurations. These differences are well below the $1\sigma$ deviation, as shown in the bottom panels of the same figure for configurations considering modes below $\sim0.8\,h\,$Mpc$^{-1}$. Departures above $\sigma$ are observed for smaller scales configurations (e.g. $\sim 1\,h\,$Mpc$^{-1}$).

One has to be cautious interpreting these results, and potentially over-weighting the differences found in the temperature.  

Despite of suppressing the effects of cosmic-variance (linked to the small volume of the simulation) by using the same DMDF at the calibration and sampling procedures (contrary to the stage at which mocks will be produced using independent DMDF, see BAM-II), the statistical description of the gas temperature is degraded by the unfair representation of high-density regions (i.e., knots, where baryon processes such as shock-heating and feedback with $T_{\rm gas}>10^{5}$K shape the small-scale signal of the temperature power-spectrum).
While this is not so dramatic for the  gas density, which appears to be homogeneously distributed on large scales (see Fig.~\ref{fig:slices_gas}), the temperature field  shows only a few prominent structures in the considered volume (see Fig.~\ref{fig:temperature_slices}). 

Nonetheless, a higher precision in the two- and three-point statistics of the temperature field could be achieved by using the full bias relation $\mathcal{P}(T_{\rm gas}|\rho_{\rm gas},n_{HI},\{\Theta\}_{\rm dm})$. This would incur in a higher memory consumption than the one considered in this study, although still far below a full hydrodynamic simulation.

Hence, we leave a study with the inclusion of non-local bias relations (through the I$^{\delta}$-web and I$^{\phi}$-web) and cosmic-variance effects on the description of the temperature field for future work, based on larger volume hydrodynamic simulations.

%In \S\ref{sec:corr} we argued, based on Fig.~\ref{fig:cross_corr_refs}, that the temperature distribution can be sensitive to the cosmic web through the invariants $I^{\phi,\delta}_{i}$. Indeed, Fig.~\ref{fig:temperature_slices} shows that using only these invariants 

%Even though the calibration performed through the bias $\mathcal{P}(T_{\rm gas}|\rho_{\rm gas},n_{HI},\delta_{\rm dm})$ aimed at a compromise between reproducing the thermodynamic properties of the gas (linking $T_{\rm gas}$, $\rho_{\rm gas}$ and $n_{HI}$), replicating the statistics of spatial distribution (through the kernel $\mathcal{K}$) and an acceptable time and memory consumption, the dependencies can be enlarged to account for anisotropies of the DMDF. 

%Indeed, the observed bias can be associated with the lack of an explicit modeling of the anisotropies of the DMDF, which, as shown in Fig.~\ref{fig:hydro_cw}, can reflect the different thermodynamic states of the gas not fully encapsulated in the simple local dependence with $\rho_{\rm gas}$, $n_{HI}$ and $\delta_{\rm dm}$.

The precision reached with the \texttt{BAM} approach already at this stage is highly competitive when compared to other methods \citep[see e.g.][]{VillaescusaNavarro2018,Ando20, Dai2020}, and in fact accurate enough for many cosmological studies of the large scale structure considering scales below  $k\sim1\,h\,$Mpc$^{-1}$.

%%%%%%%%%%%%%%%%%%%%%%%%%%%%%%%%%%%%%%%%%%%%%%%%%%%%%%%
%%%%%%%%%%%%%%%%%%%%%%%%%%%%%%%%%%%%%%%%%%%%%%%%%%%%%%%

%New section
\section{\texttt{BAM} as a machine learning approach} 
\label{sec:overfitting}

The \texttt{BAM} method represents a  physically motivated supervised machine learning algorithm in which the cost function based on the power spectrum of the respective targeted variables is minimized with non-linear and non-local isotropic kernels and anisotropic explicit bias dependencies. The kernels preserve dimensionality and are iteratively learnt from the reference simulation.

Since the dependencies are physically motivated (tidal field tensor, thermodynamical equilibrium conditions, etc.) we can save a lot of computational efforts in the machine learning process and the training set can be smaller.

As mentioned in \S\ref{sec:bias}, a non-trivial aspect of our analysis consists in addressing the potential overfitting of the multidimensional probability distribution, possibly arising when the dimensionality of the data-set (i.e. the total number of cells used in the interpolation of the DM tracer property) and the total number of bins $N_{\rm{\lbrace\Theta\rbrace}}$ are comparable. Although this will be conclusively assessed in forthcoming publications, we present here some preliminary arguments to prove that our results are not significantly affected by this issue.

Given $N_{\Theta}=10^6$ bins used to describe the total set of properties $\lbrace\Theta\rbrace$ and the total number of cells  $N_{\rm{cells}}=128^3$, the average number of cells per $\Theta$-bin is $N_{\rm{cells}}/N_{\Theta}\sim 2$, hence apparently incurring into overfitting. However, the binning adopted for the set of properties $\lbrace\Theta\rbrace$ is such that $\gtrsim 90\%$ of such bins are devoid of cells. Therefore, the calculation presented above can be misleading, as it implicitly assumes that all the $\Theta$-bins are occupied. In order to properly account for this fact, we replace $N_{\Theta}$ with an effective number of bins $N^{\rm eff}_{\Theta}$, obtained from the bins which contain at least one cell, leading to  $N_{\rm{cells}}/N^{\rm eff}_{\Theta} \sim 15$, indicating that the degree of overfitting is much lower than naively expected.

To further address this problem, we have performed two additional reconstructions of the gas density field on the same grid-size in two different scenarios, namely: (a) using I$^\delta$-web (from the dark matter) and lowering the number of bins (with respect to our fiducial value) by a factor of $4$, i.e, $N_{\rm{cells}}/N^{{\rm eff}}_{\Theta}\sim 85-90$, and (b) using local $\delta_{\rm dm}$ and modelling the bias with $N_{\Theta}=10^6$ bins, i.e. as our fiducial case. Case (a) leads to the same level of precision in the reconstruction of the power-spectrum  of the gas density field as the fiducial case, and comparably accurate bi-spectra. Case (b) yields the same level of precision in the gas density power-spectrum  reconstruction as well (even slightly improved), while it does not reach the accuracy in the bi-spectrum reconstruction achieved in the fiducial case with the same number of bins (but with the additional non-local contribution supplied by I$^{\delta}$-web). These results show that the accuracy in our reconstructions depends on the amount of physical information included in the bias modelling, and not by overfitting due to the number of bins adopted to represent the multi-dimensional tracer-bias.

What we learn from these additional studies is that the number of bins is not so crucial as one would expect since the random shuffling of cells within the \texttt{BAM} learning process breaks the summary statistics. This happens when the bias dependencies are not accurate, as in case (b). However, introducing the proper non-local bias dependencies reproduces to great accuracy the gas distribution, as in case (a).
By augmenting the training set using either larger volumes or number of reference simulations, overfitting issues can be overcome relying on the very same bias relations \citep[for the combination of simulations to overcome overfitting see][]{Balaguera2019}.

In future work, we will implement different binning strategies which can help to evaluate the risk of overfitting within the \texttt{BAM} machinery and to overcome overfitting when larger volumes simulations are required. To this end, we will soon present a new study learning from a larger volume hydrodynamic simulation (Sinigaglia et al, in prep.).

% ===============================================================================================
\section{Conclusions} \label{sec:conclusions}

This work presents an investigation of the assembly bias of baryonic quantities with respect to the dark matter field and its cosmic web characterisation.

As we have found, a natural hierarchy of bias relations emerges. The first one comes from considering the cross correlations between the baryon quantities and the dark matter.
The ionized gas density displays a fair representation of the underlying cosmic web. Nonetheless, non-local bias at small scales becomes very relevant.
While the bias learning \texttt{BAM} approach achieves within 1 \% precision in the two-point statistics, it only achieves  high accuracy in the three-point statistics (compatible in general within 1-$\sigma$) when there is an additional classification of the dependency of the ionized gas density to the invariants of the tidal density tensor in addition to the local density.
The neutral hydrogen requires the information of the ionized gas density. In a last step, the temperature can be obtained from knowing both the gas and the neutral hydrogen density. 
In this way a hierarchy of bias relations is established  which ultimately permits a practical guide to reproduce hydrodynamic simulations starting from the dark matter field.

In particular, we have applied the \texttt{BAM} method to replicate gas density, HI number density and temperature fields of a cosmological hydrodynamic simulation run on a comoving volume of $(100\,h^{-1}\rm{Mpc})^{3}$. 
We have shown to reproduce the spatial distribution of the target properties with average (throughout all probed Fourier modes ) precision of $\lesssim1\%$ in the power-spectrum  up to the Nyquist frequency ($k_{\rm{nyq}}\sim 4\, h\rm{Mpc}^{-1}$) for the gas and HI distributions, and $\sim 6\%$ for the gas temperature. These figures are translated to $5-6\%$ in the reduced bi-spectrum of gas and HI when considering configurations with triangle sides below $k\ \sim 1 h\rm{Mpc}^{-1}$. If we consider even smaller scales for the ionized gas density case  ($1 \lesssim k /( h\rm{Mpc}^{-1}) \lesssim 2.0 $) we find that a combination of large-scale and small-scale anisotropic clustering description becomes relevant. We leave an investigation of those scales for future studies. 
The bi-spectrum of the temperature field is reconstructed with average residuals of $8-10\%$.

For each of the explored gas properties, we investigated suitable bias models to reproduce the summary statistics to percent accuracy, based on local and non-local terms. We adopt the I$^{\phi}$-web classification based on the main invariants of the tidal field tensor \citep{Kitaura2020}, to describe long-range non-local terms. We account for short-range dependencies by extracting the main invariants of the tensor $\zeta_{\rm{ij}}=\partial_i\partial_j\delta$ \citep{Peacock1985,1986ApJ...304...15B}. 
These sets of invariants allow to describe the shape of the density peaks and the anisotropic clustering of structures in the cosmic web.

In our analysis, we started by reconstructing the gas density field, which displays the highest correlation with the dark matter density field. We find that the relevant information on the spatial distribution of gas at the considered scales is embedded in the short-range non-local terms, meaning that the gas is sensitive to the three-dimensional shape of the peaks of the dark matter density field, and in general to anisotropies of the short-range surrounding environment. We have then reconstructed the neutral hydrogen density based on the properties of the gas distribution to finally reconstruct the gas temperature using both dark matter and baryon properties. 
In a forthcoming paper, we will investigate the performance of this approach starting with different approximate gravity solvers based on Lagrangian perturbation theory or fast particle mesh solvers, using in the neutral hydrogen  and temperature reconstruction the quantities obtained from \texttt{BAM}.

A direct extension of our work consists in the assessment of the suitability of this technique to produce Lyman-$\alpha$ forest mock catalogs. We plan to investigate this point in a forthcoming publication, in which we will apply the same procedure presented in this paper to replicate the Lyman-$\alpha$ forest field extracted from the same reference simulation, and to study its bias with the dark matter field. Furthermore, our method can be similarly be applied to the generation of 21cm-line mock catalogs for HI intensity mapping, for large-scale structure and galaxy formation and evolution studies.

\section*{Acknowledgments}
%\acknowledgments
We acknowledge the referee for his/her comments, which have helped to improve the quality of the manuscript. We thank Marc Huertas-Company for valuable comments and insights on machine learning. FS acknowledges the support of the Erasmus+ Programme of the European Union and of the doctoral grant funded by the University of Padova and by the Italian Ministry of Education, University and Research (MIUR). FS is also grateful to the \emph{Instituto de Astrof\'{\i}sica de Canarias} (IAC) for hospitality and computing resources.
FSK and ABA acknowledge the IAC facilities and  the Spanish Ministry of Economy and Competitiveness (MINECO) under the Severo Ochoa program SEV-2015-0548, AYA2017-89891-P and CEX2019-000920-S grants. FSK also thanks the  RYC2015-18693 grant.
KN is grateful to Volker Springel for providing the original version of {\sc GADGET-3}, on which the {\sc GADGET3-Osaka} code is based on. Our numerical simulations and analyses were carried out on the XC50 systems at the Center for Computational Astrophysics (CfCA) of the National Astronomical Observatory of Japan (NAOJ), {\sc Octopus} at the Cybermedia Center, Osaka University, and {\small Oakforest-PACS} at the University of Tokyo as part of the HPCI system Research Project (hp190050, hp200041).  This work is supported in part by the JSPS KAKENHI Grant Number JP17H01111, 19H05810, 20H00180.
KN acknowledges the travel support from the Kavli IPMU, World Premier Research Center Initiative (WPI), where part of this work was conducted. 
MA thanks for the support of the Kavli IPMU fellowship and the travel support by Khee-Gan Lee. We acknowledge Cheng Zhao for making available his code to measure the bi-spectrum. We thank Marc Huertas-Company for useful discussions on ML methods.

\vspace{5cm}
%\facilities{\dots}
%\software{Example of software \citep{2013A&A...558A..33A}}

%% ANSWER TO THE REFEREE
%% END ANSWER

\bibliography{sample}{}
\bibliographystyle{aasjournal}
\end{document}